\documentstyle[12pt,epsf]{article}

\newfont{\lfig}{cmr10 scaled\magstep0} % 10 points sans serif

\newcommand{\MeV}{{\rm MeV}}
\newcommand{\GeV}{{\rm GeV}}
\newcommand{\keV}{{\rm keV}}
\newcommand{\fm}{{\rm fm}}
\renewcommand{\Im}{{\rm Im}}
\renewcommand{\Re}{{\rm Re}}

\chardef\atcode=\catcode`\@
\catcode`\@=11
\@addtoreset{equation}{section}
\renewcommand{\theequation}{\arabic{section}.\arabic{equation}}

\begin{document}

\title{Effective Lagrangian approach to vector mesons, their structure and
decays$^{*)}$}

\author{F. Klingl, N. Kaiser and W. Weise \\
Physik Department, Technische Universit\"{a}t M\"{u}nchen\\
   Institut f\"{u}r Theoretische Physik, D-85747 Garching, Germany}

\bigskip

\bigskip

\maketitle 
\begin{abstract}

          An improved update of the structure and decays of $\rho^0$,
          $\omega$ and
          $\phi$ mesons based on a chiral SU(3) Lagrangian, including anomaly
          terms is presented. We demonstrate that a consistent and
          quantitatively successful description of both pion and kaon
          electromagnetic form
          factors can be achieved. We also discuss the $e^+e^- \to \pi^+ \pi^0
          \pi^-$ cross section, the Dalitz decay 
           $\omega \to  \pi^0 \mu^+ \mu^-$ and
          aspects of $\rho^0 \omega$ and $\omega \phi$ mixing. Relations to
          previous versions of the Vector Meson Dominance model will be
          examined.

\end{abstract} 
\vspace{1.0in}

Submitted to Z.Phys.A

\vspace{.5in}
\noindent $^{*)}${\it Work supported in part by BMBF and GSI}

\newpage

\section{Introduction}

The aim of the present work is to summarize and update the physics of the
light neutral vector mesons ($\rho^0$, $\omega$ and $\phi$). We explore to what extent
their structure, their decays and their electromagnetic couplings can be
understood on the basis of an effective Lagrangian which approximates the
flavor SU(3) sector of QCD, with u-,d- and s-quarks, at low energy.

Vector mesons are well known to play a key role in the electromagnetic interaction
of hadrons. They represent the lowest "dipole" (spin-parity $J^\pi=1^-$ )
excitations of the QCD vacuum. The Vector Meson Dominance (VMD) model, first
developed in the late sixties \cite{1,2}, proved remarkably successful in the
description of a variety of hadron electromagnetic form factors and decays. 
Now that quantitative explorations of hadron structure using electromagnetic
probes are reaching a new stage, an update of VMD seems timely and appropriate.
This is also needed to guide investigations of possible changes of vector meson
properties in dense nuclear matter. The modern view is to combine VMD
and Chiral Dynamics, two principles governing low energy QCD, in a
suitably constructed effective Lagrangian \cite{3,4}. Our main purpose here is to
demonstrate that this works indeed and provides a reliable framework for
further studies. 

In the following chapters we develop the effective Lagrangian and discuss
important constraints from gauge invariance. We then examine a variety of
applications in confrontation with experiment: the pion and kaon
electromagnetic form factors,  P-wave $\pi \pi$ scattering in the isospin $I$=1
channel, specific processes such as
$e^+ e^-  \to \pi^+ \pi^0 \pi^-$ and the Dalitz decay $\omega \to \pi^0 \mu^+ \mu^-$, and
aspects of $\rho^0 \omega$- and $\omega \phi$-mixing. Special emphasis will be
given to the $q^2$-dependence of virtual photon-hadron couplings, with $q^2$ the
squared photon four-momentum. Such questions are relevant in the analysis of
dilepton production, a process actively discussed as a source of information
about hadron properties in compressed and excited matter.

\section{Vector Meson Dominance and the hadronic electromagnetic current}

Within flavor SU(3), the hadronic electromagnetic current of the Standard
Model is given in terms of the vector currents of u-, d- and s-quarks as
follows:
\begin{equation}
  \label{2.1}
  J_\mu^{em} = \frac{2}{3} \bar{u} \gamma_\mu u -\frac{1}{3} \left[
  \bar{d}\gamma_\mu d +\bar{s} \gamma_\mu s \right]\; ,
\end{equation}
where $u(x)$, $d(x)$ and $s(x)$ refer to the quark fields. This current can be
rewritten identically as
\begin{equation}
  \label{2.2}
  J_\mu^{em} = \frac{1}{\sqrt{2}} J_\mu ^{(\rho)} + \frac{1}{3 \sqrt{2}}
  J_\mu^{(\omega)} - \frac{1}{3} J_\mu ^{(\phi)}\; ,
\end{equation}
where we have introduced combinations which directly reflect the quark compositions of
the corresponding vector mesons:
\begin{eqnarray}
  \label{2.3}
J_\mu ^{(\rho)} & = & \frac{1}{\sqrt{2}} (\bar{u} \gamma_\mu u - \bar{d}
\gamma_\mu d)\; , \nonumber \\
J_\mu ^{(\omega)} & = & \frac{1}{\sqrt{2}} (\bar{u} \gamma_\mu u + \bar{d}
\gamma_\mu d)\; ,   \\
J_\mu ^{(\phi)} & = & \bar{s} \gamma_\mu s \nonumber \, .
\end{eqnarray}
In the following we identify always $\rho$ with the neutral $\rho^0$-meson and
drop the index. 
The decay of a given vector meson $V=\rho$, $\omega$, $\phi$ into an $e^+ e^-$
pair is described by the matrix element $\langle 0|J_\mu^{em}|V \rangle$ connecting a
vector meson with the QCD vacuum. Such matrix elements are commonly
written as 
\begin{equation}
  \label{2.4}
 \langle 0|J_\mu^{em}|V\rangle \,\, = -\frac{m^2_V}{g_V} \epsilon_\mu^{(V)} 
\end{equation}
where
\begin{eqnarray}
  \label{2.5}
  m_\rho & = & 768 \;\MeV\; , \nonumber \\
  m_\omega & = & 782 \;\MeV\; ,\\
 m_\phi &= &1019 \; \MeV \; \nonumber  
\end{eqnarray}
are the vector meson masses and $\epsilon_\mu^{(V)}$ is a polarization vector. The
vector meson couplings $g_V$ are determined by their $e^+e^-$
decay widths (neglecting the electron mass)
\begin{equation}
  \label{2.6}
  \Gamma (V \to e^+ e^-) = \frac{4\pi \; \alpha^2}{3} \frac{m_V}{g_V^2} \; , 
\end{equation}
with $\alpha = 1/137$. From the empirical widths one finds: 
\begin{eqnarray}
  \label{2.7}
  g_\rho &=& 5.03\; , \nonumber \\
  g_\omega &=& 17.05\; ,\\
  g_\phi &=& -12.89 \nonumber\, .
\end{eqnarray}
Note that these values are not far from the expected SU(3) ratios,
\begin{equation}
  \label{2.8}
  g_\rho : g_\omega : g_\phi = 1:3:\frac{-3}{\sqrt{2}}\, .
\end{equation}
In the original formulation by Sakurai and others \cite{1} the VMD
hypothesis was written in terms of a current-field identity,
\begin{equation}
  \label{2.9}
  J_\mu^{em}= \sum_V \frac{-m_V^2}{g_V} V_{\mu}\; ,
\end{equation}
which directly relates the hadronic electromagnetic current, at the operator
level, to the neutral vector meson fields $V_\mu = \rho_\mu$, $\omega_\mu$
and $\phi_\mu$. For later purposes it is important to recall that equation
(\ref{2.9}) strictly holds only at the vector meson poles, i.e. at timelike
photon four momentum with $q^2 = m_V^2$, although it has often been used over
wider ranges of $q^2$ without further notice. 

\section{Effective Lagrangian}

In this section we summarize the effective meson Lagrangian that will be the
basis of our calculations. Our starting point is the chiral Lagrangian \cite{5},
with pseudoscalar Goldstone boson fields as collective degrees of
freedom. Vector mesons will be introduced as massive gauge particles following
the systematic approach of Schechter et al. \cite{4,27,28}. This framework to
construct a chiral and gauge invariant Lagrangian of pseudoscalars and
vector mesons is equivalent to the ''hidden gauge symmetry'' scheme of Bando et
al. \cite{3}. (For a review, see ref.\cite{6})   
\subsection{Chiral Lagrangian including vector mesons}

Our starting point is the effective chiral Lagrangian in the flavor SU(3) sector including
u-,d- and s-quarks. Its leading term is the non-linear sigma model
\begin{equation}
  \label{3.1}
  {\cal L}_0= \frac{f_\pi^2}{4} {\rm tr}(\partial_\mu U \partial^\mu U^{ \dagger})\; ,
\end{equation}
expressed in terms of the pseudoscalar Goldstone boson fields $\Phi$:
\begin{equation}
  \label{3.2}
  U = \exp \left(\frac{i \Phi}{f_\pi} \right)
\end{equation}
where $f_\pi=92.4 \, \MeV $ is the pion decay constant. In our applications the matrix
field
\begin{equation}
 \label{3.3}
  \Phi \equiv  \sqrt{2} \left(  \begin{array}{ccc} \displaystyle \frac{\pi^0}{\sqrt{2}} &
  \pi^+ & K^+ {\rule[-3mm]{0mm}{8mm}}\\ \pi^- & \displaystyle  \frac{-\pi^0}{\sqrt{2}} & K^0{\rule[-5mm]{0mm}{10mm}}  \\
   K^-   & \overline{K^0} & 0 \end{array} \right)
\end{equation}
includes the pion and kaon fields. The octet
$\eta$-meson can easily be incorporated as well, but we will not be concerned with it in the present
paper. Similarly, the neutral vector mesons $\rho^0$, $\omega$ and $\phi$ are
represented by a matrix field
\begin{equation}
 \label{3.4}
  V_\mu  \equiv  \left(  \begin{array}{ccc} 
         \rho+\omega & 0 & 0 {\rule[-3mm]{0mm}{8mm}} \\
             0 & -\rho+\omega & 0  {\rule[-3mm]{0mm}{8mm}}\\
             0   & 0 & \sqrt{2} \phi \end{array} \right)_\mu \; ,
\end{equation}
with the field tensor 
\begin{equation}
  \label{3.5}
  G_{\mu\nu}= \partial_\mu V_\nu - \partial_\nu V_\mu\, .
\end{equation}
For most purposes in the present context, only the neutral vector mesons are
relevant, so that we can simplify the description by omitting the charged
$\rho^\pm$ and $K^*$ fields and their contribution to the (generally non-abelian) field
tensor. The coupling of the vector mesons to the pseudoscalar Goldstone bosons
is introduced through the gauge-covariant derivative
\begin{equation}
  \label{3.8}
  \partial_\mu \Phi  \to D_\mu \Phi =  \partial_\mu \Phi  + \frac{ig}{2} [\Phi,V_\mu]\; ,
\end{equation}
with the vector coupling constant $g$. 
In the following we shall work to leading order in the pseudoscalar field
$\Phi$, so that the basic Lagrangian derived in the appendix from Schechter's approach reduces to
\begin{equation}
  \label{3.11}
  {\cal L} = {\cal L}_\Phi+{\cal L}_V+{\cal L}_{V \Phi}^{(1)}+{\cal L}_{V \Phi}^{(2)}\, .
\end{equation}
${\cal L}_\Phi$ and ${\cal L}_V$ are the free field terms
\begin{eqnarray}
  \label{3.12}
{\cal L}_\Phi & = &
  \frac{1}{4} {\rm tr} \big( \partial_\mu \Phi \partial^\mu \Phi \big)
- \frac{1}{4}  {\rm tr} \big({\bf m_\Phi}^2 \Phi^2   \big)\; ,\\
 {\cal L}_V & = & -\frac{1}{8} {\rm tr} \big( G_{\mu\nu} G^{\mu\nu} \big)
+\frac{1}{4}  {\rm tr}\big({\bf m_V}^2 V_\mu V^\mu \big)\; ,
\end{eqnarray}
 with the pseudoscalar and vector meson mass matrices
\begin{equation}
  \label{3.10}
   {\bf m_\Phi}^2  \equiv   \left(  \begin{array}{ccc} 
          m_{\pi}^2 & 0 & 0 \\
             0 & m_{\pi}^2 & 0 \\
             0   & 0 & 2\,m_K^2-m_\pi^2 \end{array} \right)\; , \hspace{1cm} 
  {\bf m_V}^2  \equiv  \left(  \begin{array}{ccc} 
          m_{\rho,\omega}^2 & 0 & 0 \\
             0 & m_{\rho,\omega}^2 & 0 \\
             0   & 0 & m_\phi^2 \end{array} \right) \, .
\end{equation}
The pseudoscalar mass terms introduce explicit chiral symmetry breaking and
SU(3) flavor breaking by the pion mass $m_\pi=139.6 \; \MeV$ and the kaon
mass $m_K = 493.6 \; \MeV$ (We neglect the small mass differences between the
charged and the neutral pseudoscalar mesons). In the vector meson mass matrix we ignore, at this level, the $\rho$-$\omega$ mass difference.

The interaction parts, illustrated in Fig.1, are given by:
\begin{eqnarray}
  \label{3.14}
 {\cal L}_{V \Phi}^{(1)} & = & \frac{ig}{4}  {\rm tr} \big(V^\mu \big[ \partial_\mu
   \Phi, \Phi \big] \big)\; , \\
 {\cal L}_{V \Phi}^{(2)} & = & -\frac{g^2}{16}  {\rm tr} \big( \big[V^\mu, \Phi \big]^2 \big)\, . 
\label{3.15}
\end{eqnarray}
Eq.(\ref{3.14}) represents the standard coupling of the vector mesons to the
vector current of the pseudoscalar mesons, while (\ref{3.15}) is a
$VV\Phi \Phi$ four-point interaction required by the gauge principle leading
to eq.(\ref{3.8}).
   
\subsection{Anomalous couplings}

Apart from the "standard" interactions ${\cal L}_{V \Phi}^{(1,2)}$ there exist
terms resulting from the chiral anomaly which generates, for example, the
three-pion decay vertex $V \to 3\pi$ and the $V \to V \pi^0$ process (see
Fig.2). These mechanisms introduce two additional coupling constants, $h$ and
$g_{VV\Phi}$, in the "anomalous" interaction term
\begin{eqnarray}
  \label{3.17}
  {\cal L}_{V \Phi}^{(3)} & = &  \frac{i\,h}{4f_\pi^3}
  \epsilon^{\mu\nu\alpha\beta} {\rm tr} \big( V_\mu \partial_\nu \Phi
  \partial_\alpha  \Phi\partial_\beta\Phi\big) \nonumber \\ & + &
  \frac{g_{VVP}} {4 f_\pi} \epsilon^{\mu\nu\alpha\beta}
  {\rm tr} \big( \partial_\mu V_\nu  V_\alpha \partial_\beta \Phi \big)\; ,
\end{eqnarray}
to be added to ${\cal L}$ of eq.(\ref{3.11}). The characteristic feature of
these anomalous couplings is the appearance of the totally antisymmetric
Levi-Civita-tensor $\epsilon^{\mu\nu\alpha\beta}$ which indicates a mismatch of
natural parity assignments in these processes.

\subsection{Photon couplings}

Electromagnetic interactions of the pseudoscalar and vector mesons are
introduced in accordance with U(1) gauge invariance. The following terms are then added to eq.(\ref{3.11}):
\begin{equation}
  \label{3.18}
  {\cal L}_{em} = -\frac{1}{4}  F_{\mu \nu}F^{\mu\nu} + {\cal L}_{\gamma
  \Phi}+{\cal L}_{\gamma V}\; ,
\end{equation}
where 
\begin{equation}
  \label{3.19}
  F_{\mu\nu}=\partial_\mu A_\nu - \partial_\nu A_\mu
\end{equation}
is the electromagnetic field tensor and $A_\mu$ denotes the photon field. The
interaction terms are (with at most one photon and one vector meson)
\begin{eqnarray}
  \label{3.20}
  {\cal L}_{\gamma \Phi} &=& \frac{i e}{2} A^\mu {\rm tr}( Q [\partial_\mu \Phi,\Phi])\; , \\
\label{3.21}
{\cal L}_{\gamma V} &=& -\frac{e}{2 g_\gamma} F^{\mu \nu} {\rm tr}(Q G_{\mu \nu})\; , 
\end{eqnarray}
where 
\begin{equation}
  \label{3.22}
  Q=    \left(  \begin{array}{ccc} 
          \frac{2}{3}  & 0 & 0 \\
             0 & -\frac{1}{3} & 0 \\
             0   & 0 & -\frac{1}{3}   \end{array} \right)\, .
\end{equation}
is the quark charge matrix within flavor SU(3). We use units with
$e^2/4\pi=\alpha=1/137$.
Eq.(\ref{3.20}) represents the standard coupling of the photon to the
electromagnetic current of the pseudoscalar mesons. The photon- vector meson
coupling, eq.(\ref{3.21}), introduces a constant $g_\gamma$. Note that our ansatz
involves the photon and vector meson field tensors, as in ref.\cite{6}, so that
gauge invariance is automatic. Note also that at this stage we do
\underline{not} impose the universality condition $g_\gamma \equiv g$ which is
commonly used in the simplest version of the VMD model. In the appendix we
outline how one arrives at the photon coupling terms (\ref{3.20}, \ref{3.21})
 using the systematic approach of Schechter et al..

Finally, the anomalous vector-pseudoscalar meson interactions introduced in the
previous section imply corresponding anomalous photon couplings as illustrated
in Fig.3. They involve direct photon interactions with three pseudoscalar
mesons or with a $V\Phi$ pair:

\begin{eqnarray}
  \label{3.23}
  {\cal L}_{\gamma \Phi}^{(3)}= \frac{i c}{2 f_\pi^3} \epsilon^{\mu\nu
  \alpha\beta} A_\mu {\rm tr}(Q\partial_\nu \Phi \partial_\alpha \Phi \partial_\beta
  \Phi)\nonumber \\ 
+\frac{d}{f_\pi} \epsilon^{\mu\nu\alpha\beta} F_{\mu\nu} {\rm tr}(QV_\alpha
\partial_\beta \Phi)\, .
\end{eqnarray}
These terms are analogous to eq.(\ref{3.17}), with one vector meson replaced
by a photon. The parameter $d$ is fixed by $V \to \pi^0 \gamma$ decays in
section \ref{se42}. In addition, constraints between $c$ and $d$ are imposed by the
analysis of $\gamma \to \pi^+ \pi ^0 \pi^-$ in the chiral limit, that is for
vanishing pion mass. One finds \cite{7}

\begin{equation}
  \label{in1}
  \frac{e}{4 \pi } = c+ 12 \frac{d g}{m_\rho^2}f_\pi^2\, .
\end{equation}

 To complete this exposition, we include the vertex
which represents the "anomalous" $\pi^0 \to 2\gamma$ decay
\begin{equation}
  \label{3.24}
  {\cal L}_{\gamma \gamma \Phi}= \frac{3 e^2}{4 \pi^2 f_\pi}
  \epsilon^{\mu\nu\alpha\beta}\partial_\mu A_\nu \partial_\alpha A_\beta tr(Q^2
  \Phi)\; ,
\end{equation}
with its coefficient fixed by the chiral anomaly.
The corresponding decay width is given by

\begin{equation}
  \label{3.25}
  \Gamma(\pi^0 \to 2 \gamma) = \frac{\alpha^2 m_\pi^3}{64 \pi f_\pi^2}\, .
\end{equation}
The anomalous process $\pi \to 2 \gamma$ is not affected
by vector mesons to leading order.

\section{Vector meson decays}

\subsection{Decays into dileptons}

Vector meson decays into $e^+e^-$-pairs have already been used in section 2 in
order to fix the electromagnetic couplings $g_V$. Here we give a complete and
more precise account of $V \to l^+l^-$ decays ( where $l^\pm$ refers to $e^\pm$
or $\mu^\pm$) in the context of the effective Lagrangian
(\ref{3.18}, \ref{3.21}). The vector meson-photon interaction (\ref{3.21}) is

\begin{equation}
  \label{4.1}
  {\cal L}_{\gamma V} = - \frac{e}{2} F^{\mu\nu} \bigg( \frac{\rho_{\mu
  \nu}}{g_\rho}+\frac{\omega_{\mu \nu}}{g_\omega}+\frac{\phi_{\mu \nu}}{g_\phi}
  \bigg)\; ,
\end{equation}
where we have introduced the neutral vector meson field tensors

\begin{equation}
  \label{4.2}
 \rho_{\mu \nu}= \partial_\mu \rho_\nu-\partial_\nu \rho _\mu \; , \; \mbox{etc.}\; , 
\end{equation}
as in eq.(\ref{3.5}). The specific couplings
\begin{equation}
  \label{4.3}
  g_\rho=g_\gamma \; ,\; \; \; g_\omega= 3 g_\gamma \; ,\; \; \; g_\phi= -\frac{3}{\sqrt{2}} g_\gamma\; ,
\end{equation}
are all expressed in terms of one vector meson-photon coupling constant
$g_\gamma$. Note that the matrix element for direct conversion of a photon of
four-momentum $q$ into a vector meson in now

\begin{equation}
  \label{4.4}
  \langle V|\gamma (q)\rangle = \frac{-e\, q^2 }{g_V} \epsilon^{(\gamma)} \cdot
  \epsilon^{(\rho)}\; , \hspace{1cm} (V=\rho^0,\omega,\phi)\; ,
\end{equation}
which coincides with eq.(\ref{2.9}) at $q^2=m_V^2$ but vanishes for real
photons with $q^2=0$. The dilepton decay widths are given by

\begin{equation}
  \label{4.5}
\Gamma {(V \to l^+ l^-)} =  \left. \frac{4 \pi}{3} \, \left(
\frac{\alpha}{g_V} \right)^2
\left(1+\frac{2 \; m_l^2}{q^2} \right) \sqrt{q^2-4 m_l^2} \right|_{(q^2=m_V^2)}\; ,
\end{equation}
where $m_l$ is the lepton mass. The empirical partial widths are summarized in
table \ref{tab1}. 
The coupling constants $g_V$ deduced from these widths are listed in table \ref{tab2},
including errors.
The empirical constants evidently follow the SU(3) pattern (\ref{4.3}) with

\begin{equation}
  \label{4.6}
  g_\gamma \simeq 5.7
\end{equation}
quite well, with the exception of $g_\rho$ for which one observes a 12 \%
discrepancy.

\subsection{\label{se42} Decays into $\pi^0 \, \gamma$}

The decays $V \to \pi^0 \gamma$ can be used to fix the parameter $d$ of one of
the anomalous interaction terms in ${\cal L}^{(3)}_{\gamma \Phi}$ of
eq.(\ref{3.23}). The following expressions hold for the $\rho \to \pi^0 \gamma
$ and $\omega \to \pi^0 \gamma$ decay widths:

\begin{equation}
  \label{4.7}
  \Gamma {(\rho \to \pi^0 \gamma)}  =  \frac{1}{9} \Gamma {(\omega \to  \pi^0
  \gamma)}=  \frac{1}{24 \pi}
  \bigg(\frac{d}{f_\pi} \bigg)^2\bigg( \frac{q^2-m_\pi^2}{ \sqrt{q^2}}\bigg) ^3\; , \nonumber \\
\end{equation}
where $q^2$ equals $m_\rho^2$ or $m_\omega^2$, respectively. The measured
widths \cite{8}
\begin{eqnarray}
  \label{4.8}
 \Gamma {(\rho   \to \pi^0 \gamma)} & = & ( 119 \pm 30) \, \keV\; , \\ 
  \label{4.9}
 \Gamma {(\omega \to  \pi^0 \gamma)} & = & ( 717 \pm 42) \, \keV\; ,
\end{eqnarray}
roughly follow the SU(3) prediction of their ratio. The extracted coupling
constant is

\begin{equation}
  \label{4.10}
 d \simeq 0.01\; , 
\end{equation}
within errors of about 20\%.
Note that in our approach there is no two-step contribution $\omega \to \rho^0
\pi^0 \to \pi^0 \gamma $ to eq.(\ref{4.9}). The reason is that the
$q^2$-dependent interaction (\ref{4.1}, \ref{4.4}) which describes the $\rho^0 \to
\gamma$ conversion vanishes for a real photon.
The $\phi \to \pi^0 \gamma$ decay width vanishes as long as the $\phi$-meson is
a pure $\bar{s}s$ state. The
measured width \cite{8}

\begin{equation}
  \label{4.11}
 \Gamma (\phi   \to \pi^0 \gamma)  =  ( 5.8 \pm 0.6 ) \, \keV\; , 
\end{equation}
is, however, significantly different from zero. This is primarily due to 
$\omega \phi$-mixing. In  fact this process can be used to estimate the mixing
angle $\epsilon_{\omega \phi}$. With inclusion of both $\omega \phi$- and $\rho
\phi$-mixing, the partial decay width becomes

\begin{equation}
  \label{4.12}
 \Gamma (\phi \rightarrow \pi^0 \gamma) = \left. \frac{3}{8 \pi}
\bigg| \epsilon_{\omega \phi}+\frac{\epsilon_{\rho \phi}}{3} \bigg|^2 
\bigg(\frac{d}{f_\pi} \bigg)^2 \bigg( \frac{q^2-m_\pi^2}{ \sqrt{q^2}}\bigg)^3 
 \right|_{(q^2=m_\phi^2)}  
\end{equation}
Assuming $\epsilon_{\rho \phi }$ to be negligibly ($\epsilon_{\rho \phi } \sim
10^{-3}$), one finds, using eq.(\ref{4.11}):

\begin{equation}
  \label{4.13}
 \epsilon_{\omega \phi} \simeq 5.8 \cdot 10^{-2} \, .
\end{equation}

\subsection{\label{se43} Hadronic decays}

Table \ref{tab3} summarizes the partial widths of vector meson decays into two
or three pseudoscalar mesons. 
Note that the $\pi^+ \pi^-$ decay modes of the $\omega$ and $\phi$ are a measure of $\rho \omega$- and
$\rho \phi$-mixing, respectively. Also, the non-vanishing
$\pi^+\pi^-\pi^0$-decay width of the $\rho$-meson is due to the isospin
breaking effect of the $\rho
\omega$-mixing \cite{31}. For later discussions, observe the relatively large $\phi \to 3
\pi$ branching ratio. 
We now proceed first with an evaluation of the $\rho \to \pi^+ \pi^-$ and $\phi
\to K \overline{K}$ decays. Then we discuss the decays into 3 pions.

\subsubsection{Decays into pseudoscalar meson pairs}

These decays are described by ${\cal L}^{(1)}_{V \Phi}$ of
eq.(\ref{3.14}). Written out in explicit form, the relevant terms become

\begin{eqnarray}
  \label{4.14a}
  {\cal L}_{\rho \pi} &= &g \rho^\mu (\pi^+ \partial_\mu \pi^- - \pi^- \partial_\mu
  \pi^+)\; , \\
  \label{4.14b}
 {\cal L}_{\phi K} &= &-\frac{g}{\sqrt{2}}  \phi^\mu \bigg(K^+ \partial_\mu K^- -
 K^- \partial_\mu  K^+ + K^0 \partial_\mu \overline{K^0} - \overline{K^0} \partial_\mu K^0 \bigg)\, .
\end{eqnarray}
At tree level, the $\rho \to \pi^+ \pi^-$ and $\phi \to K^+ K^-$ decay widths
derived from eqs.(\ref{4.14a}, \ref{4.14b}) are

\begin{eqnarray}
  \label{4.15}
  \Gamma {(\rho \rightarrow \pi^+ \pi^-)} & = &  \frac{g^2}{48 \pi}
  m_\rho \bigg(1-\frac{4 m_\pi^2}{m_\rho^2} \bigg)^\frac{3}{2} \; ,\\
\label{4.16}
\Gamma {(\phi \rightarrow K^+ K^-)} & = & \frac{g^2}{96 \pi} m_\phi
  \bigg(1-\frac{4 m_{K^+}^2}{m_\phi^2} \bigg)^\frac{3}{2}\, .
\end{eqnarray}
For the $\phi \to K_L^0 K_S^0$ decay, the charged kaon mass in eq.(\ref{4.16})
is to be replaced by $m_{K^0}$. The tree-level coupling constant $g$ derived
from the empirical decay widths is given in table \ref{tab4}.
One observes that the underlying SU(3) symmetry is quite well fulfilled: values
of $g$ deduced from $\phi \to K \overline{K}$ decays differ by only about 10\% from
$g_{\rho \pi \pi} \simeq 6$. We identify $g \equiv g_{\rho \pi \pi}$ in the
following.

\subsubsection{\label{se432}Decays into three pions}

G-parity conservation implies that the only possible three-pion decay of a
vector meson in leading
order is $\omega \to 3 \pi$, and C-conjugation excludes the 3$\pi^0$ final state. Two mechanisms contribute to
this decay, both governed by ${\cal L}^{(3)}_{V\Phi}$ of eq.(\ref{3.17}). The
first one is the direct process which involves the coupling constant $h$. The
second one is the Gell-Mann, Sharp, Wagner (GSW) process \cite{9}, a two-step
mechanism in which the $\omega$ first converts into $\rho \pi$ (with coupling
constant $g_{VVP}$), followed by the decay of the virtual $\rho$ into $\pi
\pi$. The explicit expression for the $\omega \to \pi^+ \pi^0 \pi^-$ decay
width becomes

\begin{equation}
  \label{4.17}
  \Gamma ({\omega \rightarrow 3 \pi})=\left.\frac{\sqrt{q^2}}{192 \pi^3} \int \int dE_+\; dE_- \big[
\vec{p}_-^{\,2} \vec{p}_+^{\,2}-(\vec{p}_- \cdot \vec{p}_+)^2\big] \; |F|^2 \right|_{q^2=m_\omega^2}\, .
\end{equation}
Here $q$ denotes the four-momentum of the $\omega$-meson, $E_{\pm}$ and
$\vec{p}_\pm$ are the energies and momenta of the charged pions in the final
state measured in the $\omega$-meson rest frame. The energy integration is limited to $m_\pi \le E_\pm \le m_\omega-2
m_\pi$ and $2m_\pi < E_+ + E_- < m_\omega-m_\pi$ with the constraint
$\vec{p}_-^{\,2 } \vec{p}_+^{\,2}-(\vec{p}_- \cdot \vec{p}_+)^2 > 0 $ which
defines the kinematically allowed region in the $E_+ E_-$ plane. The amplitude is a sum of direct and GSW terms:
\begin{equation}
  \label{4.18}
  F=F_{dir}+F_{GSW} \, .
\end{equation}
The direct term is simply
\begin{equation}
  \label{4.19}
  F_{dir}= - \frac{3 h}{f_\pi^3}\, .
\end{equation}
The GSW term involves the intermediate $\rho$ meson. Its propagator $D_\rho
(k)= [k^2-\stackrel{\rm o}{m}_\rho^2- \Pi_\rho (k^2)]^{-1}$ includes the complex $\rho \to \pi \pi$
self-energy $\Pi_\rho(k^2)$, to be specified later, and a bare mass $\stackrel{\rm o}{m}_\rho$ such that the
physical mass is determined by $m_\rho^2=\stackrel{\rm o}{m}_\rho^2 + {\rm Re}
\Pi(k^2=m_\rho^2)$. The GSW part of the amplitude $F$ is then
\begin{equation}
  \label{4.20}
  F_{GSW} = 2 g_{VVP} \frac{g}{f_\pi} \sum_i D_\rho (q-p_i)\; ,
\end{equation}
where $p_i$ is the four-momentum of the first pion, produced in the $\omega \to
\rho \pi$ transition and the index $i$ refers to its possible charge
($i=+,0,-$).
The numerical result of the integration in eq.(\ref{4.17}) gives
\begin{equation}
  \label{4.21}
  \Gamma(\omega \to 3 \pi)/MeV=49.4 \, h^2 -27.8\, h g_{VVP}+3.98 \, g_{VVP}^2\; ,
\end{equation}
which should be equated with the empirical width $\Gamma(\omega \to 3 \pi)=
7.5 \; \MeV$.

The separate determination of $h$ and $g_{VVP}$ requires a second constraint for
which we employ the $\phi \to 3 \pi$ decay. The width of this Zweig
rule-forbidden decay is proportional to the square of the $\omega \phi$ mixing
angle $\epsilon_{\omega \phi}=5.8 \cdot 10^{-2}$ which we take from
eqs.(\ref{4.12}, \ref{4.13}).  
One finds

\begin{equation}
  \label{4.22}
  \Gamma (\phi \to 3 \pi)/MeV=|\epsilon_{\omega \phi}|^2 \big( 689 \, h^2 - 416 \, g_{VVP} h +
125\, g_{VVP}^2 \big)
\end{equation}
and this should be equal to $\Gamma (\phi \to 3 \pi)=0.68\; \MeV$.

The conditions (\ref{4.21}) and (\ref{4.22}) define two ellipses in the
$(h,g_{VVP})$ plane, as shown in Fig.4. From their intersections and the additional constraint
provided by the Dalitz plot analysis of the ratio $\Gamma(\phi \to 3
\pi)_{\rm direct} / \Gamma(\phi \to\rho \pi \to 3\pi)=(2.5 \pm0.9)/(12.9\pm0.7)$
\cite{8} we determine
\begin{eqnarray}
  \label{4.23}
  g_{VVP} & = & 1.2\; , \\
 h & = & -0.06 \, .
\end{eqnarray}
Our value for $g_{VVP}$ agrees with previous analysis \cite{4,6}, but we find a
direct $V \to 3 \pi$ coupling $h$ which is substantially larger than the value
given in ref.\cite{4}.

\section{Pion and kaon electromagnetic form factors}

The electromagnetic pion and kaon form factors have always been primary sources of
information about vector mesons. The pion form factor in the region of timelike
$q^2>0$ is dominated by the $\rho$-meson resonance, and its fine structure exhibits details
of $\rho \omega$ mixing. The kaon form factor is strongly influenced by the
presence of the $\phi$ meson, with important contributions also from $\rho$-
and $\omega$-mesons.
Our aim here is to present a unified and improved analysis of both form
 factors starting from the effective Lagrangian of section 3. The calculation
 now involves pseudoscalar meson loops and a detailed discussion of vector mesons
 self-energies.

\subsection{\label{se51} The $\rho$ meson self energy}

The $V\Phi$ interaction terms (\ref{3.14}, \ref{3.15}) of the effective Lagrangian
generate a hadronic current
\begin{equation}
  \label{5.1}
  J_{had}^\mu = i \left( \pi^{+} \, \partial^\mu \pi^{-} - \pi^{-}\:
  \partial^\mu \pi^{+} + 2i  g\, \pi^+ \pi^-  \rho^\mu \right)\, .
\end{equation}
Its coupling to the $\rho$ meson field produces the
self-energy shown in Fig.5. The resulting self-energy tensor from one pion loop
diagrams is
\begin{equation}
  \label{5.2}
  \Pi^{\mu \nu}(q) = -i g^2 \int \frac{d^4p}{(2 \pi)^4}
  \frac{1}{p^2-m_\pi^2+i\epsilon} \left[ \frac{(2p-q)^\mu
  (2p-q)^\nu}{(p-q)^2-m_\pi^2+i \epsilon} -2 g^{\mu \nu} \right]\; ,
\end{equation}
where $q$ is the four-momentum carried by the $\rho$ meson. The intermediate
pion four-momenta are $p$ and $p-q$, respectively. The full $\rho$-meson
propagator is
\begin{equation}
  \label{5.3}
  i\, D_{\rho}^{\mu\nu}(q) = \frac{-i}{q^2-\stackrel{\rm
  o}{m}_\rho^2-\Pi_\rho (q^2)} \bigg[ g^{\mu\nu} - \frac{
q^\mu q^\nu}{q^2} \bigg]+\frac{i}{\stackrel{\rm o }{m}_\rho^2} \frac{q^\mu q^\nu}{q^2}\, .
\end{equation}
with
\begin{equation}
  \label{5.4}
  \Pi_\rho (q^2) =  \frac{1}{3} g_{\mu\nu} \Pi^{\mu\nu} (q).
\end{equation}
We have introduced the bare $\rho$ meson mass $\stackrel{\rm o }{m}_\rho$ so that
its physical mass is given by
\begin{equation}
  \label{5.5}
  m_\rho^2 = \stackrel{\rm o }{m}_\rho^2+ \Re \Pi_\rho (q^2=m_\rho^2)\, .
\end{equation}
The $\rho \to \pi \pi$ decay width at resonance is
\begin{equation}
  \label{5.6}
  \Gamma (\rho \to \pi \pi) = - \Im  \Pi_\rho (q^2=m_\rho^2) / m_\rho \, .
\end{equation}
The imaginary part of $\Pi_\rho$ is easily calculated using standard Cutkosky
rules \cite{10}, with the result
\begin{equation}
  \label{5.7}
  \Im \, \Pi_\rho (q^2) = - \frac{g^2}{48 \pi \sqrt{q^2}} (q^2-4
  m_\pi^2)^{\frac{3}{2}} \Theta ( q^2 - 4 m_\pi^2 ) \; .
\end{equation}

\subsection{Regularization}

The loop integral (\ref{5.2}) is divergent and needs to be regularized. Whereas
the authors of ref.\cite{11} have employed the Pauli-Villars regularization
method 
\cite{12}, we prefer to use a (twice) subtracted dispersion relation for the
self-energy:
\begin{equation}
  \label{5.8}
  \Pi_\rho (q^2) = c_0+c_1 \;q^2+\frac{q^4}{\pi} \int_{4 m_\pi^2}^\infty  ds \; \frac{\Im\,\Pi_\rho (s)}{s^2
\; (s-q^2-i \epsilon)}\, .
\end{equation}
Inserting eq.(\ref{5.7}) one finds for the real part of $\Pi_\rho$:
\begin{equation}
  \label{5.9}
  \Re \: \Pi_\rho (q^2)  =  c_0+c_1 \;q^2-\frac{g^2}{24\pi^2} \left[ q^2 \;
  ({\cal G} (q^2,m_\pi^2) +\frac{4}{3}) -4 m_\pi^2 \right]\, .
\end{equation}
where
\begin{equation}
  \label{5.10}
  {\cal G}(q^2,m^2) = \left\{ \begin{array}{r@{\quad:\quad}l} 
\left(\frac{4m^2}{q^2}-1 \right)^{\frac{3}{2}} \arcsin{\frac{\sqrt{q^2}}{2\: m}} & 0<q^2<4m^2 \\ -\frac{1}{2} 
\left(1-\frac{4m^2}{q^2} \right) ^{\frac{3}{2}} \ln{\left[
\frac{1+\sqrt{1-\frac{4m^2 }{q^2}}}{1-\sqrt{1-\frac{4m^2 }{q^2}}} \right] } &
4m^2<q^2 \; {\rm or}\; q^2<0 \end{array}
\right. \, . 
\end{equation}
The subtraction constants $c_0$ and $c_1$ need still to be determined. First we
note that $c_0$ should vanish. In order to see this we examine the
corresponding pion-loop self-energy of the photon which appears in the pion
form factor and has exactly the same form as eq.(\ref{5.2}) except that $g^2$
is replaced by $e^2$. In particular, the subtraction constants in the
corresponding dispersion relation (\ref{5.8}) are the same in both cases. The requirement
that the photon stays massless at $q^2=0$ then immediately implies $c_0=0$.

To determine $c_1$ we take a more detailed look at the $\rho$ mass
renormalization procedure. Expanding the transverse part of the
$\rho$-propagator (\ref{5.3}) around the physical mass $m_\rho$ we have
\begin{equation}
  \label{5.11}
  \frac{1}{q^2-\stackrel{\rm o}{m}^2_\rho-\Pi_\rho (q^2)} \simeq
\frac{Z}{(q^2-m_\rho^2)-i Z \: \Im_\rho \:\Pi (q^2)}\; ,
\end{equation}
with the renormalization constant
\begin{equation}
  \label{5.12}
  Z = \left(1- \frac{d}{dq^2} \Re \:\Pi_\rho (q^2)|_{(q^2=m_\rho^2)} \right)^{-1}\, .
\end{equation}
One can introduce a ''bare'' coupling constant $\stackrel{ \rm o}{g}$ which is
related to the physical coupling constant by $\stackrel{ \rm o}{g}=
Z^{-\frac{1}{2}} g$. We now demand that the effect of the real part of the
self-energy is completely absorbed by shifting the bare mass $\stackrel{ \rm
o}{m}_\rho$ to its physical value $m_\rho$, so that $Z=1$ and the coupling
constant $g$ remains unaltered. The resulting condition
\begin{equation}
  \label{5.13}
  \frac{d}{dq^2} \Re \Pi_\rho |_{q^2=m_\rho^2}=0
\end{equation}
fixes the subtraction constant $c_1$. We introduce $x \equiv
1-4m_\pi^2/m_\rho^2$ and find
\begin{equation}
  \label{5.14}
  c_1 = \frac{g^2}{24 \pi^2} \left[ \frac{4}{3}-\frac{1}{2} \left( x+(1+\frac{2
  m_\pi^2}{m_\rho^2}) \sqrt{x} \ln{\frac{1+\sqrt{x}}{1-\sqrt{x}}} \right)
  \right] = - 0.118\, . 
\end{equation}
The coupling constant $g$ has already been determined by comparison with the
$\rho \to \pi^+ \pi^-$ decay width (see table \ref{tab4}):
\begin{equation}
  \label{5.15}
  g=6.05\, .
\end{equation}
The only remaining parameter is the ''bare'' mass $\stackrel{\rm o}{m}_\rho$
that the $\rho$ meson would have if its coupling to the $\pi \pi$ continuum
were turned off ($g \to 0$).

\subsection{P-wave isovector $\pi \pi$ phase shift}

We can fix $\stackrel{\rm o}{m}_\rho$ by the requirement that the P-wave
$\pi\pi$ scattering phase shift in the isospin $I=1$ channel be optimally
reproduced \cite{26}. The partial wave amplitude in this channel (with squared center of
mass energy $s$) becomes
\begin{equation}
t_1^1 (s)= \sqrt{\frac{s}{s-4 m_\pi^2}} e^{i \delta^1_1} 
\sin{\delta^1_1 (s)} = -\frac{g^2}{48 \pi} \frac{s -4m_\pi^2}{s- \stackrel{\rm o}{m}_\rho^2-\Pi_\rho (s)}\; ,
  \label{5.16}
\end{equation} 
so that
\begin{equation}
  \label{5.17}
  \cot \delta^1_1 (s)  =  \frac{-s+ \stackrel{\rm o}{m}_\rho^2+\Re \: \Pi_\rho
  (s)}{\Im \:\Pi_\rho (s)}\, . 
\end{equation}
Choosing
\begin{equation}
  \label{5.18}
  \stackrel{\rm o}{m}_\rho = 0.81 \; \GeV \; ,
\end{equation}
we obtain the fit shown in Fig.6 which is evidently quite
satisfactory. (Similar results are obtained with Pauli-Villars regularization
for which the bare mass turns out to be somewhat larger, $\stackrel{\rm
o}{m}_\rho = 0.83 \;\GeV$, together with a Pauli-Villars cutoff parameter
$2\Lambda_{PV}= 0.92 \;\GeV$).

At this point it is important to note that the mass shift induced by the $\rho
\to \pi \pi$ self-energy is small:
\begin{equation}
  \label{5.19}
  \frac{\stackrel{\rm o}{m}_\rho -m_\rho}{m_\rho} \simeq 5 \% \, . 
\end{equation}
This justifies, a posteriori, our calculation of the self-energy to leading
order in $g^2$.

\subsection{\label{se54} Pion form factor}

The electromagnetic pion form factor is defined by the matrix element
\begin{equation}
  \label{5.20}
  \langle \pi^\pm(k')|J_\mu^{em}(0)|\pi^\pm(k)\rangle = \pm (k+k')_\mu F_\pi (q^2)\; ,
\end{equation}
with $q^2=(k-k')^2$. The leading order, tree level terms contributing to $F_\pi
(q^2)$ are obtained from ${\cal L}_{\gamma \Phi}$ and  ${\cal L}_{\gamma V}$ of
eqs.(\ref{3.20}, \ref{3.21}) together with ${\cal L}^{(1)}_{V\Phi}$ of
eq.(\ref{3.14}) and shown in Fig.7(a,b). Their contribution is
\begin{equation}
  \label{5.21}
  F_\pi^{(0)} (q^2)  = 1-  \frac{g}{ \stackrel{ \rm o }{g}_\rho}  \frac{q^2}{q^2-\stackrel{ \rm o }{m}_\rho^2+i\: \epsilon}\, .
\end{equation}
Here we have introduced the ''bare'' $\gamma \rho$ coupling constant
$\stackrel{ \rm o }{g}_\rho$ which, unlike the $\rho \pi \pi$ coupling constant
$g$, is renormalized by pion loops. Introducing the $\rho \to \pi
\pi$ self-energy we have
\begin{eqnarray}
  \label{5.22}
  F_\pi (q^2) & = & F_\pi^{(0)}(q^2) \left( 1+ i
\Pi_\rho (q^2) \frac{- i}{q^2 -\stackrel{ \rm o }{m}_\rho^2} \right. \nonumber
\\ && \left. \hspace{1 cm}
+i\Pi_\rho (q^2) \frac{- i}{q^2 -\stackrel{ \rm o }{m}_\rho^2} i\Pi_\rho (q^2) \frac{- i}{q^2
-\stackrel{ \rm o }{m}_\rho^2}+... \right)\, .
\end{eqnarray}
When summed up to all orders this becomes
\begin{equation}
  \label{5.23}
   F_\pi (q^2) = \frac{(1- \frac{g}{\stackrel{ \rm o }{g}_\rho})q^2-\stackrel{ \rm o }{m}_\rho^2}{q^2-\stackrel{ \rm o }{m}_\rho^2-\Pi_\rho(q^2)}\, .
\end{equation}
Note that not only the $\rho$ meson propagator, but also the $\gamma \rho$
coupling is modified by pion loops (see Fig. 7(c)):
\begin{equation}
  \label{5.24}
  \langle \gamma(q)|\rho \rangle= -\frac{e\, q^2}{\stackrel{ \rm o }{g}_\rho}+\frac{  e
\Pi_\rho (q^2)}{g}\, .
\end{equation}
We can absorb this vertex correction in the definition of an effective $\gamma \rho$ coupling strength:
\begin{equation}
  \label{5.25}
   \frac{1}{g_\rho(q^2)} \equiv \frac{1}{\stackrel{ \rm o
   }{g}_\rho}-\frac{\Pi_\rho(q^2)}{g \, q^2}\; ,
\end{equation}
so that the pion form factor becomes
\begin{equation}
  \label{5.26}
   F_\pi (q^2) =  1-  \frac{g}{g_{\rho}(q^2)}  \frac{q^2}{q^2-\stackrel{ \rm o }{m}_\rho^2-\Pi_\rho (q^2)}\, .
\end{equation}
The constant $g_\rho$ determined from the $\rho \to e^+ e^-$ and $\rho \to
\mu^+ \mu^-$ decay widths should be compared with $\Re \, 
g_\rho(q^2=m_\rho^2)$. Using $\Re \, g_\rho=5.0$ from table \ref{tab2} we find
\begin{equation}
  \label{5.27}
  \stackrel{ \rm o }{g}_\rho=5.44\; ,
\end{equation}
which now agrees better with the SU(3) prediction if one uses $g_\omega=17.0$
(see table \ref{tab2}).

Finally we include $\rho \omega$-mixing which introduces an additional factor
in $F_\pi(q^2)$:
\begin{equation}
  \label{5.28}
   F_\pi  (q^2)  = \left( 1-  \frac{g}{g_{\rho}(q^2)}
   \frac{q^2}{q^2-\stackrel{ \rm o }{m}_\rho^2-\Pi_{\rho} (q^2)}\right) \left(1+
\frac{g_{\rho}(q^2)}{g_{\omega}}\frac{ z_{\rho
\omega}}{q^2-m_{\omega}^2+i m_\omega \Gamma_\omega} \right).\nonumber \\
\end{equation}
with the $\omega$ meson width $\Gamma_\omega=8.4 \; \MeV$ and the mixing parameter
$z_{\rho \omega}=-4.52 \cdot 10^{-3} \, \GeV^2$ \cite{13}.

The resulting pion form factor in the region of timelike $q^2$ is shown in
Fig.8 together with data \cite{14}. The agreement is evidently very
satisfactory. The ultimate fine tuning has been achieved with $g_\rho
(q^2=m_\rho^2)=4.93$.
One should note that the present description gives a substantial improvement as
compared to the standard VMD result with constant $\gamma \rho$ coupling
$\langle \gamma |\rho \rangle = -e \stackrel{ \rm o}{m}_\rho^2/g_\rho$
which leads to : 
\begin{equation}
  \label{5.29}
  F_\pi^{VMD_0} (q^2) = \frac{g}{g_\rho}
\frac{- \stackrel{ \rm o }{m}_\rho^2}{q^2-\stackrel{ \rm o }{m}_\rho^2-\Pi_\rho (q^2)}\, .
\end{equation}
In this case the universality condition $g=g_\rho$ is mandatory in order to
make sure that $F_\pi^{VMD}(q^2=0)=1$. The corresponding optimal result for
$F_\pi$ including $\rho \omega$-mixing is shown as the dashed curve labeled
VMD$_0$ in Fig.8.

The form factor in the spacelike region $q^2 <0$ is presented in Fig.9 and
compared with data from high energy pion-electron scattering \cite{15,31}. Our
approach with $q^2$-dependent $\gamma \rho$ coupling proves once again to be
significantly better than the standard (VMD$_0$) result. The squared pion
charge radius becomes
\begin{equation}
  \label{5.30}
  \langle r_\pi^2 \rangle  = 6\; \frac{d\, F_\pi {(q^2)}}{d q^2} \bigg|_{q^2=0} =
\frac{6}{\stackrel{ \rm o }{m}_\rho^2} \left(
\frac{g}{\stackrel{ \rm o }{g}_\rho}+\frac{d \Re \Pi_\rho (q^2)}{d q^2}
\bigg|_{q^2=0} \right)\, .
\end{equation}
Using the dispersion relation (\ref{5.8}) this can be rewritten
\begin{equation}
  \label{5.31}
  \langle r_\pi^2 \rangle = \frac{6}{\stackrel{ \rm o }{m}_\rho^2} \left(
\frac{g}{\stackrel{ \rm o }{g}_\rho}-c_1 \right) = 0.44 \; \fm^2\; ,
\end{equation}
with the subtraction constant $c_1=-0.118$ as before. Eq.(\ref{5.30}) is seen
to be in perfect agreement with the empirical value $\langle r_\pi^2 \rangle =
(0.44 \pm 0.01) {\rm fm}^2$ \cite{16}, whereas the standard VMD result $\langle
r_\pi^2 \rangle =6/m_\rho^2= 0.40  \; {\rm fm}^2$ is too small.

It is also instructive to examine the chiral limit ($m_\pi \to 0$) within our
framework. In this limit the subtraction constant has the leading behavior
\begin{equation}
  \label{5.32}
  c_1=- \frac{g^2}{48 \pi^2} \ln{\frac{m_\rho^2}{m_\pi^2}}\; ,
\end{equation}
showing that the pion radius (\ref{5.30}) diverges logarithmically as the pion
becomes soft. Using the KSFR relation $\stackrel{ \rm o
}{m}_\rho^2=2g^2f_\pi^2$\cite{17} we find
\begin{equation}
  \label{5.33}
  <r_\pi^2> =  \left(\frac{1}{4 \pi \: f_\pi} \right)^2
  \ln{\frac{m_\rho^2}{m_\pi^2}} + {\rm non \,\,\, singular \,\,\,terms}\, .
\end{equation}
which is exactly the result obtained in chiral perturbation theory \cite{29}. This
confirms that the present approach, including the subtraction procedure to
handle divergent pion loops, is in accordance with the rules implied by chiral
symmetry.

\subsection{Kaon form factor}

The electromagnetic form factor of the charged kaon is defined by
\begin{equation}
  \label{5.34}
  \langle K^\pm (k')|J_\mu^{em}(0)| K^\pm (k) \rangle = \pm   (k+k')_\mu F_K (q^2)\, .
\end{equation}
Its prominent feature is the presence of the $\phi$ meson just above $K\overline{K}$
threshold. One can therefore expect that the leading behavior of $F_K(q^2)$ is
obtained by just transcribing the previous formalism developed for $F_\pi(q^2)$,
replacing the $\rho$ meson by the $\phi$ meson and the pion loop by the kaon
loop. We find
\begin{equation}
  \label{5.35}
   F_K^{(\phi)} (q^2) =  1+  \frac{g}{\sqrt{2}g_{\phi}(q^2)}  \frac{q^2}{q^2-\stackrel{ \rm o }{m}_\phi^2-\Pi_\phi (q^2)}\; ,
\end{equation}
where we have used the SU(3) relation $g_{\phi K^+K^-}=-g/\sqrt{2}$. The
$\phi$ self-energy has contributions from $\phi \to K \overline{K}$ and $\phi \to
\pi^+ \pi^0 \pi^-$ couplings:
\begin{equation}
  \label{5.36}
  \Pi_\phi = \Pi_{\phi \to K^+ K^-} + \Pi_{\phi \to K^0_L K^0_S}+ \Pi_{\phi \to
  3 \pi}\; ,
\end{equation}
while the photon coupling of the $\phi$ meson, including renormalization by the
charged kaon loop, is
\begin{equation}
  \label{5.37}
  \frac{1}{g_\phi(q^2)} = \frac{1}{\stackrel{ \rm o}{g}_\phi}+\frac{\sqrt{2}}{g}
  \frac{\Pi_{\phi \to K^+ K^-}(q^2)}{q^2}\, .
\end{equation}
Note that we have $ \Re \, g_\phi (q^2=m_\phi^2) = -12.9$.
For the $K \overline{K}$ self-energies one finds
\begin{eqnarray}
  \label{5.38}
  \Re \Pi_{\phi \to K^+K^-} (q^2)&  = &c_K \, q^2 -\frac{g^2}{48 \pi^2} q^2 \left[ {\cal G}
 (q,m_{K^+}^2)- \frac{4 m_{K^+}^2}{q^2}\right]\; , \\
\Im \;\Pi_{\phi \to K^+K^-} (q^2) &  = &- \frac{g^2}{96\pi} q^2 \; \Theta (q^2-4m_{K^+}^2)\left( 1 - \frac{4m_{K^+}^2}{q^2}
 \right) ^\frac{3}{2} \; , 
\end{eqnarray}
where ${\cal G}$ is defined in eq.(\ref{5.10}) and $c_K$ is a subtraction
constant. For the real and imaginary parts of $\Pi_{\phi \to K^0_L K^0_S}$ the
same expressions hold with the charged kaon mass $m_{K^+}$ replaced by
$m_{K^0}$. The subtraction constant is determined by a condition analogous to
eq.(\ref{5.13}):
\begin{equation}
  \label{5.40}
  \frac{d}{dq^2} \Re \Pi_{\phi \to K \overline{K}} |_{q=m_\phi^2}=0\; ,
\end{equation}
which gives $c_K \simeq 0.11$.
The imaginary part of the $3\pi$ self-energy is determined by the $\phi \to
\pi^+ \pi^0 \pi^-$ decay width:
\begin{equation}
  \label{5.41}
  \Im \Pi_{\phi \to 3 \pi}= - \sqrt{q^2} \Gamma_{\phi \to \pi^+ \pi^0
  \pi^-}(q^2)\; ,
\end{equation}
where the $q^2$ dependence includes the $3\pi$ phase space (see also section \ref{se432}). The real part $\Re \Pi_{\phi \to 3 \pi}$ cannot be calculated in a
reliable and unambiguous way. Using e.g. reasonable cutoffs it turns out to be
negligible small.
We propose simply to absorb its effects by adjusting the bare mass parameter
$\stackrel{ \rm o}{m}_\phi$, so that the shift between $\stackrel{ \rm
o}{m}_\phi$ and the physical mass $m_\phi$ is generated entirely by the kaon
loops.
The bare values of the $\phi$ mass and $\phi$-photon coupling constant obtained
in this way are
\begin{equation}
  \label{5.42}
  \stackrel{ \rm o}{m}_\phi \;\simeq 0.91 \GeV, \hspace{1cm} \stackrel{\rm
  o}{g}_\phi \; \simeq -10\, .
\end{equation}
Note that the relative mass shift $(m_\phi \: -\stackrel{ \rm
o}{m}_\phi)/m_\phi$ induced by the real part of the $\phi \to K \overline{K}$
self-energy does not exceed 10\%. The bare coupling constant is slightly, but
not much smaller than the SU(3) prediction $\stackrel{ \rm o}{g}_\phi \; \simeq
-12$ based on $g_\omega=17$.

Our description of the kaon form factor is not yet complete. Contributions from
the $\rho$ and $\omega$ mesons must also be included. We use SU(3) relations to
determine their couplings to $K^+K^-$pairs: $g_{\rho K^+ K^-}=g_{\omega K^+
K^-}=g/2$. The improved kaon form factor is then:
\begin{eqnarray}
  \label{5.43}
   F_{K}(q^2) &= &1+
  \frac{g}{\sqrt{2} g_{\phi} (q^2)} \frac{q^2}{q^2
  -\stackrel{ \rm o}{m}_\phi^2-\Pi_\phi(q^2)}\nonumber \\
    &-&\frac{g}{2g_{\rho}(q^2)} \frac{q^2}{q^2-
    \stackrel{ \rm o}{m}_\rho^2-\Pi_\rho(q^2)} \nonumber \\
    &-&\frac{g}{2g_\omega(q^2)} \frac{q^2}{q^2-
    \stackrel{ \rm o}{m}_\omega^2-\Pi_\omega(q^2)} \, .
\end{eqnarray}
The $\rho$ meson self-energy now incorporates both $\rho \to \pi^+ \pi^-$ and
$\rho \to K^+ K^-$ loop contributions, and the $\rho$-photon coupling.
(We mention in passing that corrections to the pion formfactor from $K^+ K^-$
loops are small and can safely be neglected there). The primary correction in the present context comes from the photon coupling to a $\pi^+ \pi^-$ loop which
converts into $K^+ K^-$ via the intermediate $\rho$ meson. For an intermediate
$\omega$-meson, this mechanism does not exist as a leading effect. We can
therefore ignore the $q^2$ dependence of $g_\omega(q^2)$ and simply use
$g_\omega =17$. The bare masses are $\stackrel{ \rm o}{m}_\rho = 0.81 \; \GeV$ as
before (the mass shift due to the $\rho \to K^+K^-$ coupling is almost
negligible) and $\stackrel{ \rm o}{m}_\omega \: =0.80 \; \GeV \simeq \, \stackrel{ \rm
o}{m}_\rho$.
The predicted charged kaon form factor for timelike $q^2$ is shown in
Fig.10. It exhibits the $\phi$ meson as the dominant structure. Fig.11 shows
the form factor at spacelike $q^2<0$. Here it becomes evident that the $\phi$
alone is not enough to reproduce the slope of the formfactor at small
$q^2$. The $\rho$ and $\omega$ meson contributions to $F_K(q^2)$ are an important
part of the mean square radius of the charged kaon,
\begin{eqnarray}
  \label{5.44}
\langle r_{K^\pm}^2 \rangle & = & 6 \frac{d F_{K}}{d q^2} \bigg|_{q^2=0}
\nonumber \\
\nonumber &= & 6 
\left[ \bigg(\frac{g}{\stackrel{ \rm o}{g}_\rho}-c_1 \bigg)\frac{1}{2 \stackrel{ \rm
o}{m}_\rho^2} +\frac{g}{2 g_\omega \stackrel{ \rm o}{m}_\omega^2}-
\left(\frac{g}{ \sqrt{2} \stackrel{ \rm o}{g}_\phi}+c_K \right)\frac{1}{\stackrel{ \rm o}{m}_\phi^2 }
\right] \nonumber \\ &\simeq& 0.40\; \fm^2\, .
\end{eqnarray}
It agrees quite well with the empirical value \cite{18}
\begin{equation}
  \label{5.45}
  \langle r_{K^\pm}^2 \rangle = (0.34 \pm 0.05)\, \fm^2\; ,
\end{equation}
given that the $\rho K \overline{K}$ and $\omega K \overline{K}$ coupling constants can
only be estimated through SU(3) symmetry.
 
The neutral kaon form factor is obtained from eq.(\ref{5.44}) replacing the
charged $K^+K^-$ couplings by $K^0 \overline{K}^{\, 0}$ couplings at the appropriate
places. This introduces a relative change of sign for the $\rho$ with respect
to the $\omega$ and $\phi$ contributions since $g_{\rho K^0 \overline{K}^{\, 0}}=-g_{\rho K^+ K^-}=-g/2$, whereas $g_{\omega K^0
\overline{K^0}}=g_{\omega K^+ K^-}=g/2$ and $g_{\phi K^0 \overline{K^0}}=g_{\phi K^+
K^-}=-g/\sqrt{2}$. The slope of the neutral kaon form factor at $q^2=0$ now
becomes:
\begin{eqnarray}
  \label{5.46}
\langle r_{K^0}^2 \rangle & = & 6 \frac{d F_{K^0}}{d q^2} \bigg|_{q^2=0}
\nonumber \\
\nonumber &= & 6 
\left[ \bigg(-\frac{g}{\stackrel{ \rm o}{g}_\rho}+c_1 \bigg)\frac{1}{2 \stackrel{ \rm
o}{m}_\rho^2} +\frac{g}{2 g_\omega \stackrel{ \rm o}{m}_\omega^2}-
\left(\frac{g}{\sqrt{2}\stackrel{ \rm o}{g}_\phi}+c_K \right)\frac{1}{\stackrel{ \rm o}{m}_\phi^2 }
\right] \nonumber \\ &\simeq& -0.074 \; \fm^2\, .
\end{eqnarray}
In the limit $m_V \to \infty$ the result
\begin{equation}
  \label{5.i}
  \langle r_{K^0}^2 \rangle = \frac{1}{16 \pi^2 f_\pi^2} \ln{\frac{m_\pi}{m_K}}
  = -0.036
\end{equation}
from chiral perturbation theory is obtained.
Although there is a subtle cancelation involved in getting this result, it is
remarkably consistent with the empirical value \cite{19}
\begin{equation}
  \label{5.47}
   \langle r_{K^0}^2 \rangle = (-0.054 \pm 0.026) \, \fm^2\, .
\end{equation}
The scarce existing data of the $K^0$ electromagnetic form factor in the timelike region
\cite{20,25} are also well reproduced.
In summary, we find that our effective Lagrangian approach, carried to one-loop
order, gives a consistent and quantitatively satisfactory picture of both pion and
kaon form factors.

\section{Further test cases}

\subsection{The $e^+e^- \to \pi^+ \pi^0 \pi^-$ reaction}

As a further test we briefly examine the three-pion spectrum produced in
$e^+e^-$ annihilation. This process dominantly involves the $\omega$ meson and
its decay into $\pi^+ \pi^0 \pi^-$. It is also a sensitive probe of $\omega
\phi$ mixing since the $\phi$ signal clearly shows up in the $e^+ e^- \to \pi^+
\pi^0 \pi^-$cross section. The input for this calculation has already been
prepared in section \ref{se432} where the $3\pi$ decays of the $\omega$ and $\phi$
have been analyzed in order to fix the ''anomalous'' couplings $g_{VVP} \simeq 1.2$
and $h \simeq -0.06$ in the effective Lagrangian (see eq.(23)). Furthermore we
use the $\omega \phi$ mixing parameter $\epsilon_{\omega \phi} \simeq 0.058$ as
given in section \ref{se42}.

The $e^+e^- \to \pi^+ \pi^0 \pi^-$  total cross section can be written as a function
of the squared c.m. energy $s=q^2$  as follows:
\begin{eqnarray}
  \label{5.48}
 \sigma_{e^+ e^- \rightarrow \pi^+\pi^0 \pi^-} (q^2) &=& \\\nonumber \left(
 \frac{4 \pi \,
\alpha}{g_\omega} \right)^2  |\Im \, \Pi_{\omega\rightarrow 3 \pi}(q^2)|&& \!
\! \! \!  \left| \frac{q^2}{q^2-\stackrel{ \rm o}{m}_\omega^2-\Pi_{\omega} (q^2)}
-\epsilon_{\omega \phi} \frac{g_\omega}{g_\phi}   \frac{q^2}{q^2-\stackrel{ \rm o}{m}_\phi^2-\Pi_{\phi} (q^2)}\right|^2 \, .
\end{eqnarray}
Here the $\omega$ and $\phi$ self-energies, $\Pi_\omega$ and $\Pi_\phi$ include
the $3 \pi$ and $K \overline{K}$ channels as described in section before. In
particular, the imaginary part of $\Pi_{\omega \to 3 \pi}$ is related to the
$q^2$-dependent $\omega \to \pi^+ \pi^0 \pi^-$ decay width:
\begin{equation}
  \label{5.49}
  \Im \Pi_{\omega \to 3 \pi} (q^2) =-\sqrt{q^2} \Gamma_{\omega \to 3 \pi} (q^2)\, .
\end{equation}
In eq.(\ref{5.48}) we have ignored the $q^2$-dependence of the $\omega$-and
$\phi$-photon couplings, $g_\omega (q^2)$ and  $g_\phi (q^2)$, and simply used
$g_\omega=17$ and $g_\phi=-12$, which is sufficiently accurate in the present
context. The bare $\omega$- and $\phi$-masses are $\stackrel{ \rm
o}{m}_\omega=0.80 \GeV$, $ \stackrel{ \rm o}{m}_\phi = 0.91 \GeV$ as before.

The calculated cross section (\ref{5.48}) is presented in Fig.12 together with
the available data. The comparison is quite satisfactory, not only around the
$\omega$- and $\phi$-resonance structures, but also for the $3 \pi$ continuum
between them. The interference pattern around the $\phi$ resonance is very
sensitive to the precise value of the constant $\epsilon_{\omega \phi} g_\omega /
g_\phi$. With $g_\omega/g_\phi = -\sqrt{2}$ determined by SU(3) (which is our
input), the only remaining parameter is $\epsilon_{\omega \phi}$ which in turn
is constrained by the $\phi \to \pi^0 \gamma$ decay width (see section
\ref{se42} ). One should note that $\epsilon_{\omega \phi}$ is only determined
up to a phase $e^{i \delta}$. A small $\delta \ne 0$ might help to improve the
comparison with the cross section above the $\phi$-resonance. Together with the kaon form factor, the overall consistency in describing
such a variety of data is non-trivial.

\subsection{The Dalitz decay $\omega \to \pi^0 \mu^+ \mu^-$}

We finally turn to a case which is again sensitive to $\omega \phi$ mixing
and, in addition, touches some basic questions about the $q^2$ dependence of
photon-vector meson couplings. Consider the Dalitz decay process $\omega \to
\pi^0 \mu^+ \mu^-$ which, unlike the $\omega \to \pi^0 \gamma$ decay discussed
previously, involves a virtual photon whose squared four-momentum $q^2$ is
monitored by the center-of-mass energy of the produced $\mu^+ \mu^-$ pair. In
our approach, the $\omega \to \pi^0 \gamma$ decay with a real photon cannot be
described by the sequence $\omega \to \rho \pi^0 \to \gamma \pi^0$ since the
$\rho \gamma$ coupling vanishes at $q^2=0$. On the other hand, the two-step
mechanism $\omega \to \rho^0 \pi^0 \to \mu^+ \mu^- \pi^0$ does contribute to the
production of $\mu^+ \mu^-$ pairs, with $q^2 >0$.
We study the Dalitz decay form factor defined by
\begin{equation}
  \label{7.1}
  F(q^2) = \frac{\langle \omega | \pi^0 \gamma^*(q^2) \rangle}{\langle \omega | \pi^0 \gamma(q^2=0) \rangle}\, .
\end{equation}
We include both the one-step process $\omega \to \pi^0 \gamma^*$ described by
${\cal L}^{(3)}_{V\Phi}$ of eq.(\ref{3.17}) and subsequent $\rho$ to photon
conversion (see Fig.13). One finds
\begin{equation}
  \label{7.2}
  F(q^2)=  1- \frac{e\; g_{VVP}}{6\,d \, g_{\rho}
  (q^2)}\frac{q^2}{q^2-\stackrel{\rm o}{m}_\rho^2 -
\Pi_{\rho} (q^2)} \; ,
\end{equation}
where the $\rho$ meson self-energy $\Pi_\rho (q^2)$, its bare mass
$\stackrel{\rm o}{m}_\rho$ and the renormalized $\rho \gamma$-coupling
$g_\rho(q^2)$ have been discussed previously in section \ref{se51}-\ref{se54}. Universality
in the sense of the ''old'' VMD model would imply
\begin{equation}
  \label{7.3}
  \frac{d}{e}=\frac{g_{VVP}}{6 g_\gamma}
\end{equation}
with $g_\gamma \equiv g_\rho$, so that in this case the form factor reduces to
\begin{equation}
  \label{7.4}
  F^{VMD_0}(q^2)= \frac{-\stackrel{\rm o}{m}_\rho^2}{q^2-\stackrel{\rm o}{m}_\rho^2- \Pi_\rho (q^2)}\; ,
\end{equation}
which happens to be identical to the pion form factor, $F_\pi^{VMD_0}(q^2)$ of
eq.(\ref{5.29}) in the same limit. By comparison with existing data for
$F(q^2)$ (see Fig.14) it is clear, however, that simple universality does not
seem to work.

The results of the full calculation are shown in Fig.14 with inclusion of
$\omega \phi$-mixing for different choices of the mixing parameter
$\epsilon_{\omega \phi}$. As seen in section \ref{se43} the value of $g_{VVP}$ is determined through the $\omega \phi$ mixing. As a consequence our result is very sensitive to the mixing parameter $\epsilon_{\omega \phi}$. For the preferred value $\epsilon_{\omega \phi} = 5.8
\cdot 10^{-2}$ the result almost coincides with that of the simple vector meson
dominance model. One would have to go to smaller mixing parameters for further
improvement. We want to stress that the coupling $g_{VVP}$ is determined at the
physical $\omega$ meson mass. This means that any form factor multiplied to this
coupling constant cannot reduce the apparent discrepancy. Improved data especially at small $q^2$ would be needed to reach more precise conclusions.

\section{Concluding remarks}

We have demonstrated that the structure and decays of the light neutral vector mesons
($\rho^0$, $\omega$ and $\phi$) are well described using a chiral SU(3) effective
Lagrangian including anomaly terms. In particular, the $q^2$-dependent gauge invariant coupling of the vector mesons to photon gives excellent
results for the pion and kaon form factors, both at timelike and spacelike
$q^2$. Loop corrections produce shifts from the bare to the physical
vector meson masses which turn out to be small, typically less than 10\% of
their physical masses.
Only the form factor of the Dalitz decay $\omega \to \pi^0 \mu^+ \mu^-$ still
leaves some open questions.

\renewcommand{\theequation}{\Alph{section}.\arabic{equation}}
\appendix
\section{Chiral framework for VMD}
Our effective Lagrangian (\ref{3.11}) for the interacting pseudoscalar-, vector
meson and photon system derives from the systematic approach of Schechter et
al. \cite{4,27,28} which is based on chiral symmetry and gauge invariance. If we
work up to quadratic order in $\Phi$ the Lagrangian of eq.(6) and (12) in
ref.\cite{28} reads
\begin{eqnarray}
  \label{ap1}
  {\cal L} &=& -\frac{1}{4} F_{\mu\nu} F^{\mu \nu} -\frac{1}{8} {\rm tr}(G_{\mu \nu} G^{\mu
  \nu})+\frac{1}{4} {\rm tr} (\partial_\mu \Phi \partial^\mu \Phi)-\frac{1}{4} {\rm tr}(
  {\bf m_\Phi} \Phi^2) \nonumber \\ & +& \frac{1}{4} m_V^2 {\rm tr} ( V_\mu V^\mu)-\frac{e m_V^2}{g_\gamma}
  A_\mu {\rm tr}( Q V^\mu ) + m_V^2 \left( \frac{e}{g_\gamma} \right)^2 A_\mu^2
  {\rm tr}(Q^2)\nonumber \\
&+&i\frac{g_\gamma k}{8} {\rm tr} (V_\mu [\partial^\mu \Phi, \Phi])+i\frac{e}{2}\left( 1-\frac{k}{2} \right) A_\mu {\rm tr} (Q [\partial^\mu \Phi, \Phi]) 
\end{eqnarray}
with the constraints
\begin{eqnarray}
  \label{ap2}
  g=g_\gamma \frac{k}{2} \; ,\\
  \label{ap3}
  k g_\gamma^2 f_\pi^2 = m_V^2. 
\end{eqnarray}
We add the mass term for the pseudoscalar mesons, neglect for simplicity the anomaly
sector and consider the vector meson masses to be equal (SU(3) symmetry). Our
notation for fields and coupling constants is connected to the one of Schechter (subscript S) in the following way:
\begin{equation}
  \label{ap4}
  \Phi_S = \frac{1}{\sqrt{2}} \Phi\,, \hspace{0.5cm} V_S=\frac{1}{\sqrt{2}}V\,,
  \hspace{0.5cm} \nonumber F_{\pi,S } =\sqrt{2} f_\pi \,,\hspace{0.5cm} g_{\rho \pi
  \pi,S}=\sqrt{2} g \,, \hspace{0.5cm}  g_S= \frac{g_\gamma}{\sqrt{2}}\, .
\end{equation}
If one considers only the $\rho$-meson, one can easily check that the Lagrangian of eq.(\ref{ap1}) is equivalent to a
alternative method of Bando et al \cite{3}, called hidden-symmetry approach. In the hidden-symmetry approach the
$\rho$-meson is a dynamical gauge boson of a hidden local symmetry assumed to
be created by
quantum effects. 

We can transform eq.(\ref{ap1}) to our $q^2$-dependent model of
eqs.(\ref{3.12}-\ref{3.14}) by using \cite{30}
\begin{eqnarray}
  V_\mu &\rightarrow &V_\mu + \frac{2 e}{g_\gamma}  A_\mu Q, \nonumber \\
  A_\mu &\rightarrow &\sqrt{1- \frac{4e^2}{3 g_\gamma^2}} A_\mu, \nonumber \\
 e &\rightarrow &\frac{e}{\sqrt{1-\frac{4e^2}{3g_\gamma^2}}}\, .
\label{ap5}
\end{eqnarray}
If $k=2$ ($a=2$ in the Bando et al. model) we get the standard VMD model with universality ($g=g_\gamma$) and
the KSFR-relation \cite{17} $m_V=\sqrt{2} g f_\pi$. On the
other hand if we take the values of $g_\gamma$ and $g$ as determined from the
pion form factor (or $\rho \to e^+ e^-$ decay width) and $\pi \pi$ scattering
($\rho \to \pi^+ \pi^-$) decay we find
\begin{eqnarray}
  \label{ap6}
k & = & \frac{\stackrel{\rm o}{m}_\rho^2}{g_\gamma^2 f_\pi^2} =2.42 \\
\tilde{k} & \equiv  & 2 \frac{g}{g_\gamma}=2.16  \, .
\end{eqnarray}
In Schechter's approach $\tilde{k}$ and $k$ have to be identical
(see eqs.(\ref{ap2}, \ref{ap3})). We note however that this 
constraint $k=\tilde{k}$ can be released simply by adding a non minimal term (higher order in
derivatives) proportional to $tr \big[G_{\mu\nu}(\xi B^{\mu\nu}_R  \xi^\dagger +\xi^\dagger B^{\mu\nu}_L
 \xi  ) \big] $ with $\xi^2=U$ to the Lagrangian of Schechter \cite{28}. This term in invariant under 
 $U(3)_L \times U(3)_R$ and for electromagnetism with $B^{\mu\nu}_L=B^{\mu\nu}_R=e Q
 F_{\mu\nu}$ it gives an additional interaction of the form added to eq.(\ref{ap1}):
\begin{equation}
  \label{Ba3}
\delta{\cal L}_{V \gamma}  =  \frac{e(k-\tilde{k})}{4 g} F_{\mu\nu} {\rm tr}
(G^{\mu\nu}(\xi Q \xi^\dagger + \xi^\dagger Q \xi )) = \frac{e(k-\tilde{k})}{2 g} F_{\mu\nu} {\rm tr}
\big(G^{\mu\nu}Q \big)+\dots
\end{equation}
The additional term allows us to have $k$ and $\tilde{k}$
 different from each other without violating chiral symmetry. After performing
 the transformation \ref{ap5} we end up with the form of the effective
 Lagrangian given in the main text. The non minimal term in eq.(\ref{Ba3}) is
 mandatory to have $g$ and $g_\gamma$ a priori
 unconstrained. 

A different way of looking at this problem has been advocated by Schechter in pointing out that the vector
meson and U(1)-gauge field in eq.(\ref{ap1}) are not the physical ones.  Since there is a 
mixing term in the Lagrangian one has to diagonalize to mass eigenstates via
the transformation
\begin{eqnarray}
  \label{ba7}
 A^\mu_PQ &=& \left(1+\frac{4e^2}{g_\gamma^2}\right)^{-\frac{1}{2}} \;\left( A^\mu Q + \frac{2e}{g_\gamma} V^\mu \right) \\
V^\mu_P &=& \left(1+\frac{4e^2}{g_\gamma^2}\right)^{-\frac{1}{2}} \;\left(V^\mu
-\frac{2e}{g_\gamma} A^\mu Q\right) \, .
\end{eqnarray}
to the physical fields (subscript P).
Applying this transformation to the interaction terms in the Lagrangian one sees that the physical
photon now couples directly to the pion in the form $ e A_\mu (\pi^+
\partial_\mu \pi^- -\pi^- \partial \pi^+)$.
Obviously, after the diagonalization there is no direct interaction anymore between the
physical photon and the physical vector meson.

Now the question arises how does the $\rho$-meson enter the pion form
factor. Schechter pointed out that only the U(1) gauge boson couples to the
electron (lepton) by gauge invariance. After diagonalization the U(1) gauge
field gets however an
admixture of the physical $\rho_P$-meson. The latter now directly interacts
with the electron. This new effect does not influence the low energy lepton-lepton scattering (around
$q^2=0$), because the effective propagator of the U(1) gauge field is 
\begin{equation}
  \label{ba9}
  \frac{1}{q^2}\left( 1+\frac{\tilde{k}}{2} \frac{q^2}{q^2-m_\rho^2}\right)\, .
\end{equation}
The pion form factor on the other hand becomes
\begin{equation}
  \label{ba12}
  F_\pi (q^2)= \left( 1+\frac{\tilde{k}}{2} \frac{q^2}{q^2-m_\rho^2}\right)
\end{equation}
which is same result as in the $q^2$-dependent model (eq.(\ref{5.21}) in
 leading order (without pion loops).

We have seen that we can derive our $q^2$-dependent model from Lagrangian of
Schechter. Electromagnetisms is uniquely introduced by demanding gauge
invariance for the non linear sigma model extended by the vector meson
nonet. After adding the non minimal term eq.(\ref{Ba3}) there is no need for
universality and we can choose $g$ and $g_\gamma$ independent from each
other as done here in our $q^2$-dependent model.

\newpage

\begin{table}
\begin{center}
\begin{tabular}{|c||c|c|c|} \hline
  $ \Gamma_{V \rightarrow l^+ l^-} $ & $\rho$ & $\omega $ & $\phi$ \\ 
  \hline $e^+ e^-$ & 6.77 $\pm$ 0.32 & 0.60 $\pm$0.02 & 1.37 $\pm$0.03
  \\  $\mu^+ \mu^-$ & 6.94 $\pm$0.42 &$ < $1.5 & 1.10 $\pm$0.15 \\ \hline
\end{tabular}
\caption{ \label{tab1} Partial width (in keV) for vector meson decays into
lepton pairs (from \protect{\cite{8}})}
\end{center}
\end{table}

\begin{table}
\begin{center}
\begin{tabular}{|c||c|c|c|} \hline
  & $g_\rho$ & $g_\omega $ & $g_\phi$ \\\hline
from $V \to e^+ e^-$ & 5.03  &  17.05 &    -12.89      \\ 
from $V \to \mu^+ \mu^- $ & 4.96 & $>$ 10.7 & -14.37     \\ \hline 
SU(3) with $g_\gamma$=5.66 & 5.66 & 17.0 & -12.02  \\ \hline
\end{tabular}
\caption{ \label{tab2} Vector meson-photon coupling constants determined from
$V \to e^+ e^-,\mu^+\mu^-$ decays at $q^2=m_V^2$. The lower line gives the
constants according to eq.(\protect{\ref{4.3}}), with $g_\omega = 3 g_\gamma$ fixed as input.}
\end{center}
\end{table}

\begin{table}
\begin{center}
\begin{tabular}{|c||c|c|c|} \hline
   & $ \rho $ & $\omega $ & $\phi $\\ \hline \hline
$\Gamma_{\pi^+ \pi^-} / \Gamma_{tot}$ &  $\sim $ 100 \% & 2.2 $\pm 0.3 \cdot
10^{-2} $&  $8 \pm 5 \cdot 10^{-5} $ \\ 
$\Gamma_{K^+ K^-} / \Gamma_{tot}$ &   -  & - & $0.491 \pm 0.009$ \\ 
$\Gamma_{K^0 \overline{K^0}} / \Gamma_{tot}$ &   -  & -  &  $0.343 \pm 0.007$ \\ 
$\Gamma_{\pi^+\pi^-\pi^0}/\Gamma_{tot} $ &  $ <1.2 \cdot 10^{-4}$  & $0.888 \pm
0.006$& $0.153 \pm 0.006$ \\ \hline
$\Gamma_{tot} [MeV]$ & $150.9 \pm 3.0$  & $8.4 \pm 0.1$ & $4.43 \pm 0.06$  \\ \hline
\end{tabular}
\caption{ \label{tab3} Empirical decay widths of vector mesons into two or three
pseudoscalar mesons (from \protect{\cite{8}}).}
\end{center}
\end{table}

\begin{table}
\begin{center}
\begin{tabular}{|c||c|c|c|} \hline
decay & $\rho \to \pi^+ \pi^- $& $\phi \to K^+ K^- $&$\phi \to K^0_L K^0_S$  \\ \hline
exp. width [MeV]& $150.9 \pm 3.0 $ & $2.175 \pm 0.039$ &$1.519 \pm 0.031 $\\ 
$g$   & 6.05 & 6.46 & 6.67   \\ \hline 
\end{tabular}
\caption{ \label{tab4} Empirical widths of vector mesons decays into two
pseudoscalar mesons and extracted coupling constant $g$.}
\end{center}
\end{table}

\clearpage

{\large \bf Figure Captions:}

\vspace{0.5 cm}

\begin{figure}
\begin{center}  
\epsfysize9cm \leavevmode\epsfbox{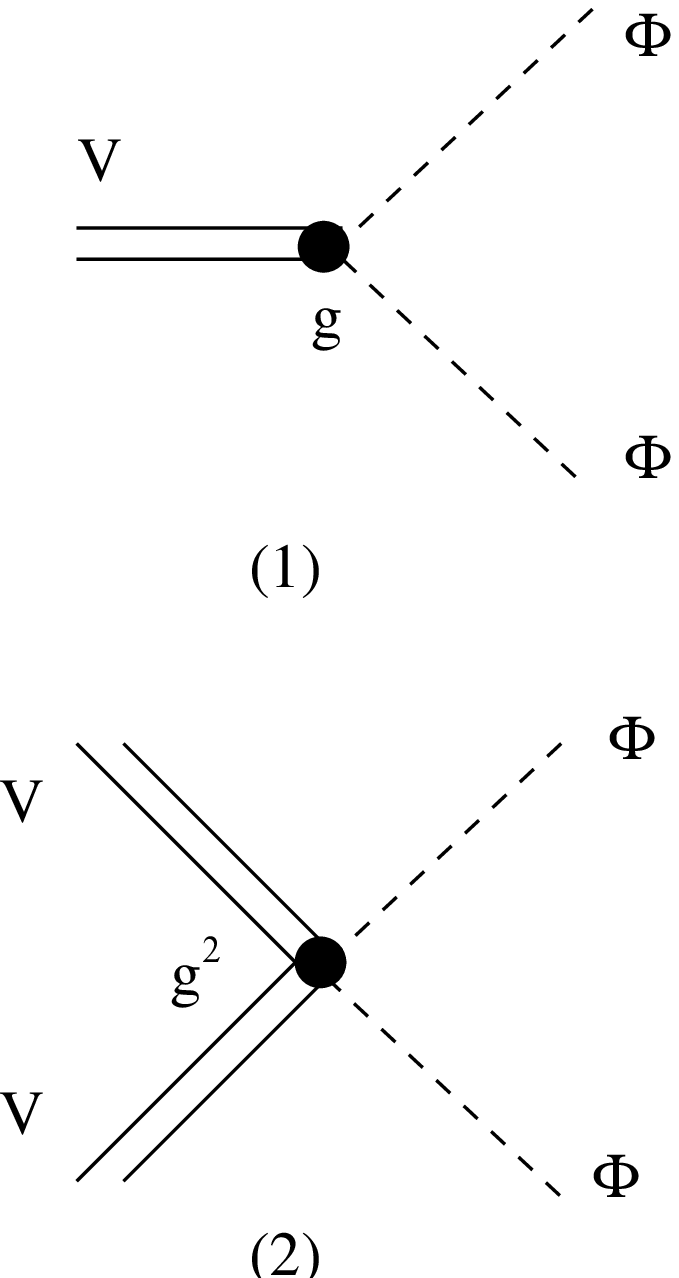}
\caption{\label{fig.1}}
\end{center} 
\end{figure}

Figure \ref{fig.1}: Three- and four point vertices describing interactions
between vector mesons and pseudoscalar mesons, as given by the
Lagrangians $ { \cal L}^{(1,2)}_{ V \Phi } $ of eqs.(\ref{3.14}, \ref{3.15}). 

\vspace{0.5 cm}

\begin{figure}
 \begin{center}  
\epsfysize9cm \leavevmode\epsfbox{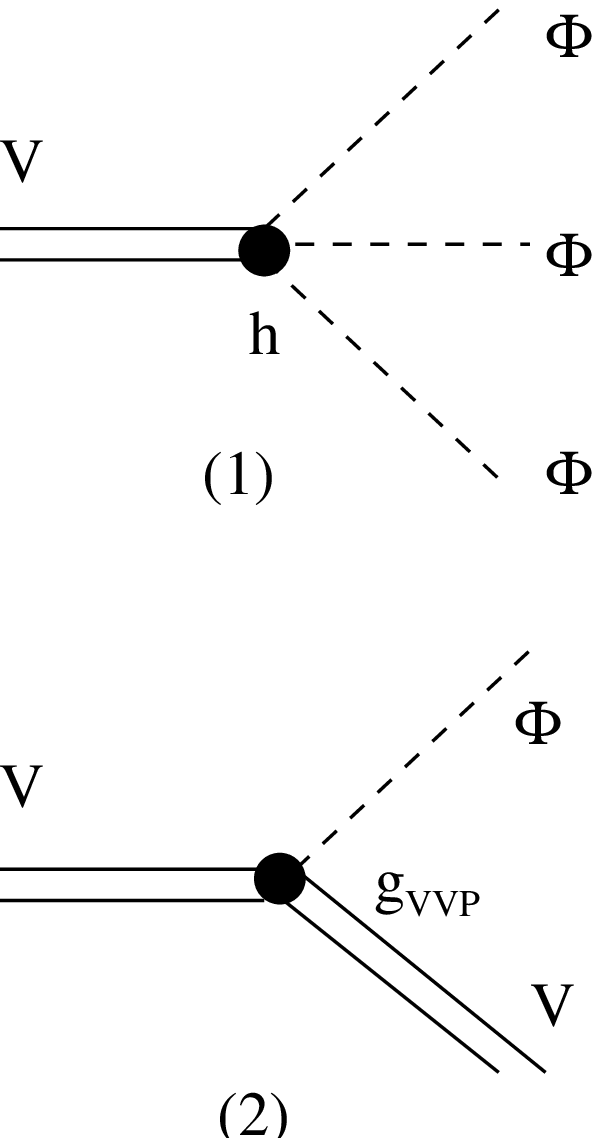}
\caption{\label{fig.2}}
\end{center} 
\end{figure}

Figure \ref{fig.2}: ''Anomalous'' couplings involving: (1) one vector and
three pseudoscalars; (2) two vector mesons and one pseudoscalar.

\vspace{0.5 cm}

\begin{figure}
 \begin{center}  
\epsfysize9cm \leavevmode\epsfbox{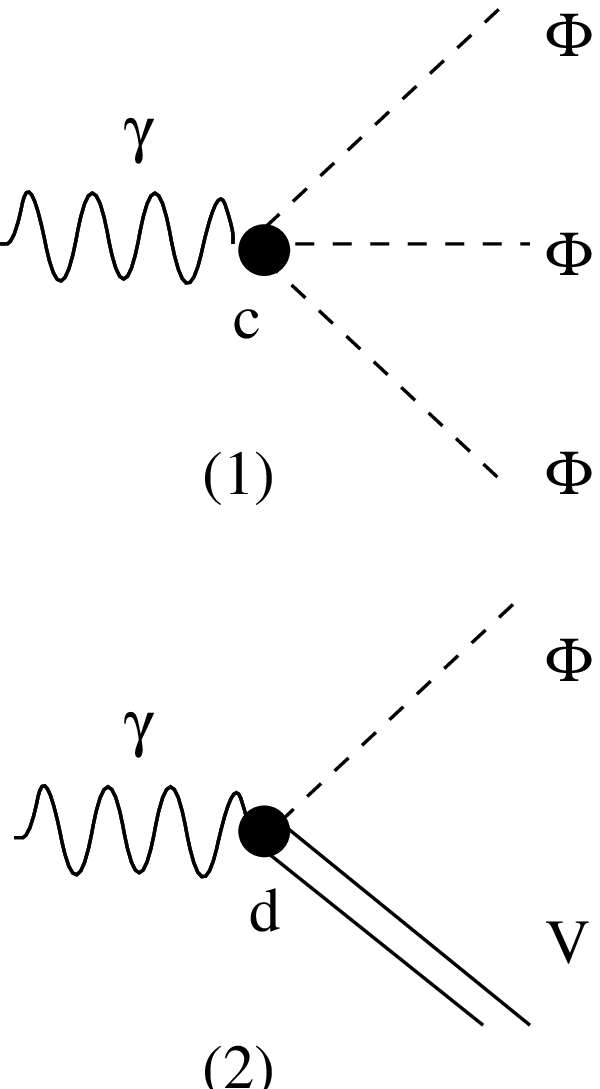}
\caption{\label{fig.3}} 
 \end{center} 
\end{figure}

Figure \ref{fig.3}: ''Anomalous'' photon couplings: (1) to three
pseudoscalar mesons; (2) to a vector meson and a pseudoscalar meson. 

\vspace{0.5 cm}

\begin{figure}
\begin{center}
\epsfysize9cm \leavevmode\epsfbox{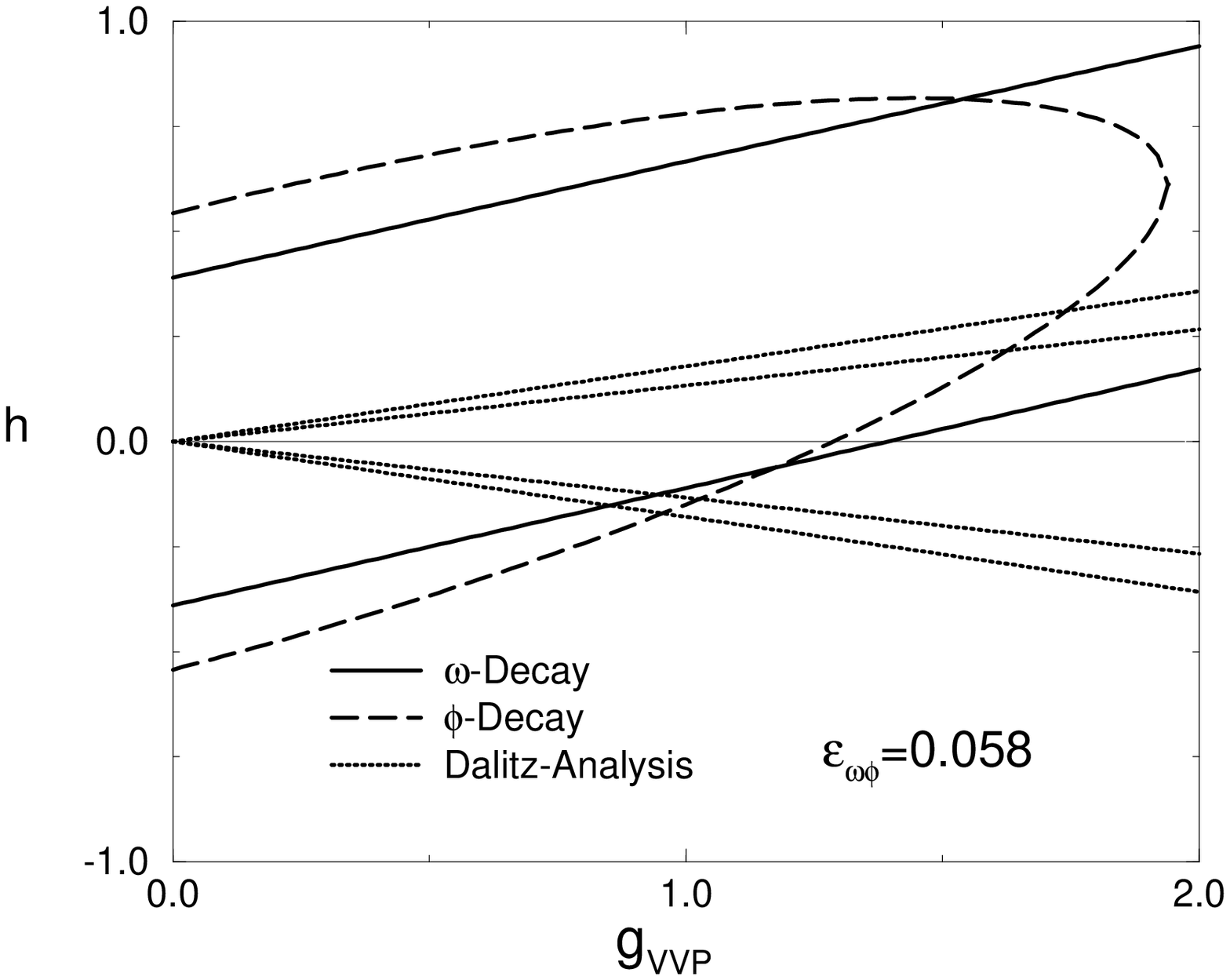}
\caption{\label{fig.4}}
\end{center}
\end{figure}

Figure \ref{fig.4}: Determination of couplings constants $g_{VVP}$ and $h$
from $3 \pi$ decays of the $\omega$- and $\phi$-meson. The solid and dashed
curves correspond to eqs.(\ref{4.21}) and (\ref{4.22}), the latter with $\epsilon=5.8
\cdot 10^{-2}$. The dotted curves represent constraints from the Dalitz plot
analysis with $\Gamma (\phi \to 3 \pi)_{\rm direct}/\Gamma (\phi \to \rho \pi
\to 3 \pi)=0.18$ and 0.10, respectively.

\vspace{0.5 cm}

\begin{figure}
 \begin{center}  
\epsfysize9cm \leavevmode\epsfbox{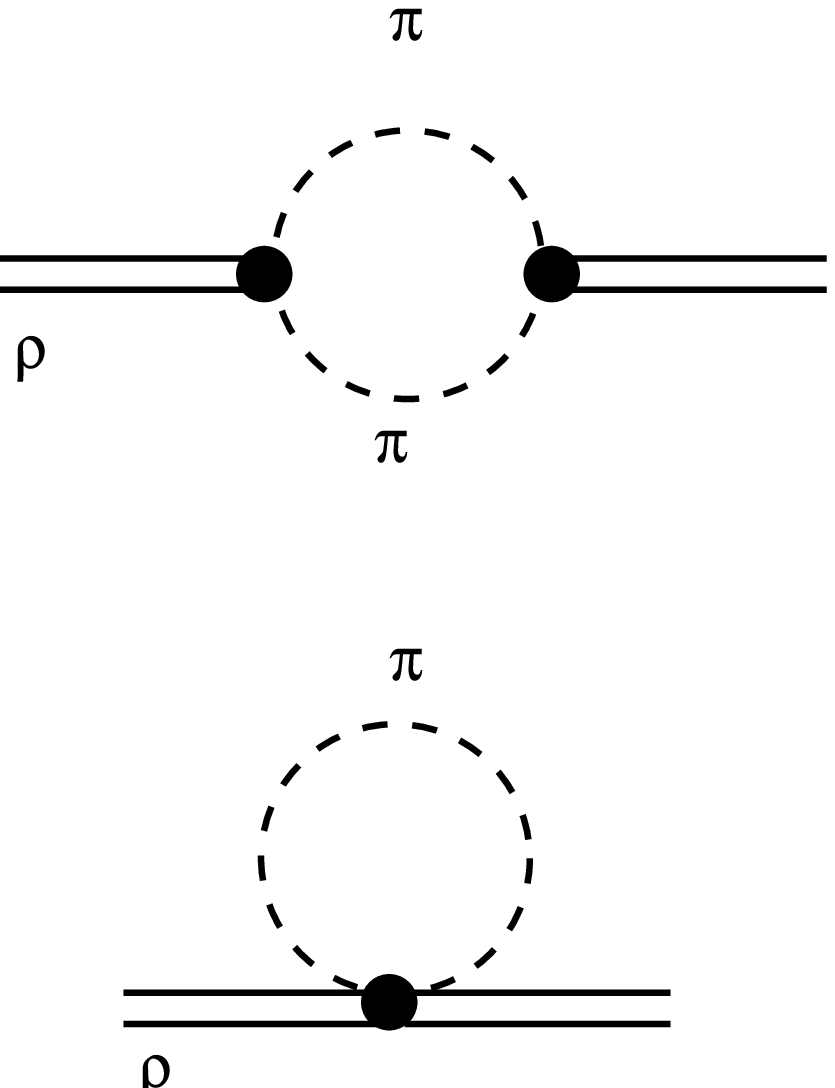}
\caption{\label{fig.5}}
\end{center}
\end{figure}

Figure \ref{fig.5}: Self-energy of the $\rho$-meson

\vspace{0.5 cm}

\begin{figure}
\begin{center} 
\epsfysize9cm  \leavevmode\epsfbox{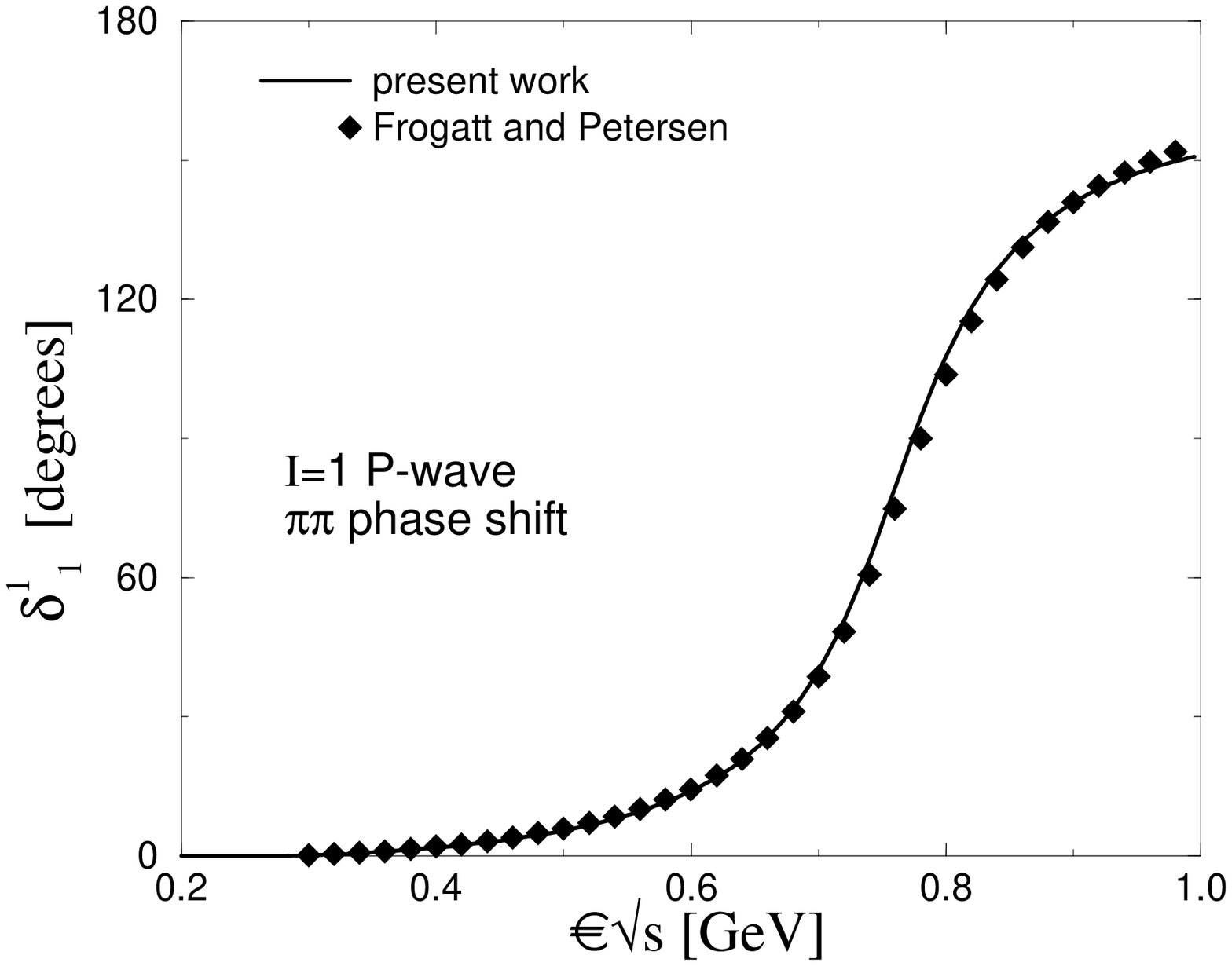}
\caption{\label{fig.6}}
\end{center} 
\end{figure}

Figure \ref{fig.6}: Isovector P-wave $\pi \pi$ scattering phase shift as
obtained from eq.(\ref{5.17}) using the subtraction constant of
eq.(\ref{5.14}) and
the bare $\rho$-meson mass $\stackrel{\rm o}{m}_\rho$=809 MeV. The phase shift
analysis is from ref.\cite{21}.

\vspace{0.5 cm}

\begin{figure}
\begin{center}
\epsfysize8cm \leavevmode\epsfbox{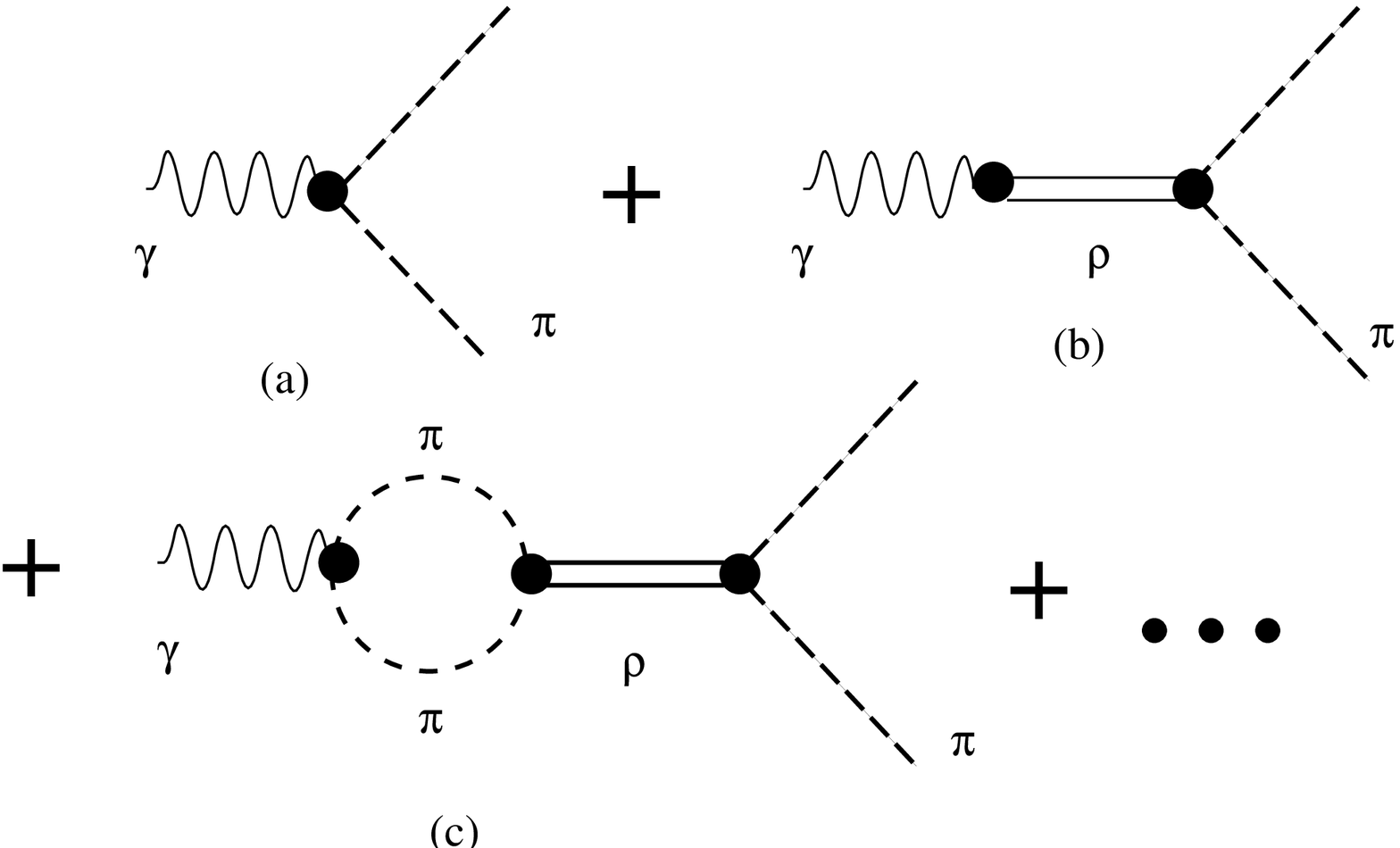}
\end{center}
\caption{\label{fig.7}}
\end{figure}

Figure \ref{fig.7}: Contributions to the pion form factor: (a) direct
coupling, (b) coupling through intermediate $\rho$-meson, (c) pion loop vertex
corrections to $\gamma \rho$ coupling.

\vspace{0.5 cm}

\begin{figure}
\begin{center}
\epsfysize9cm \leavevmode\epsfbox{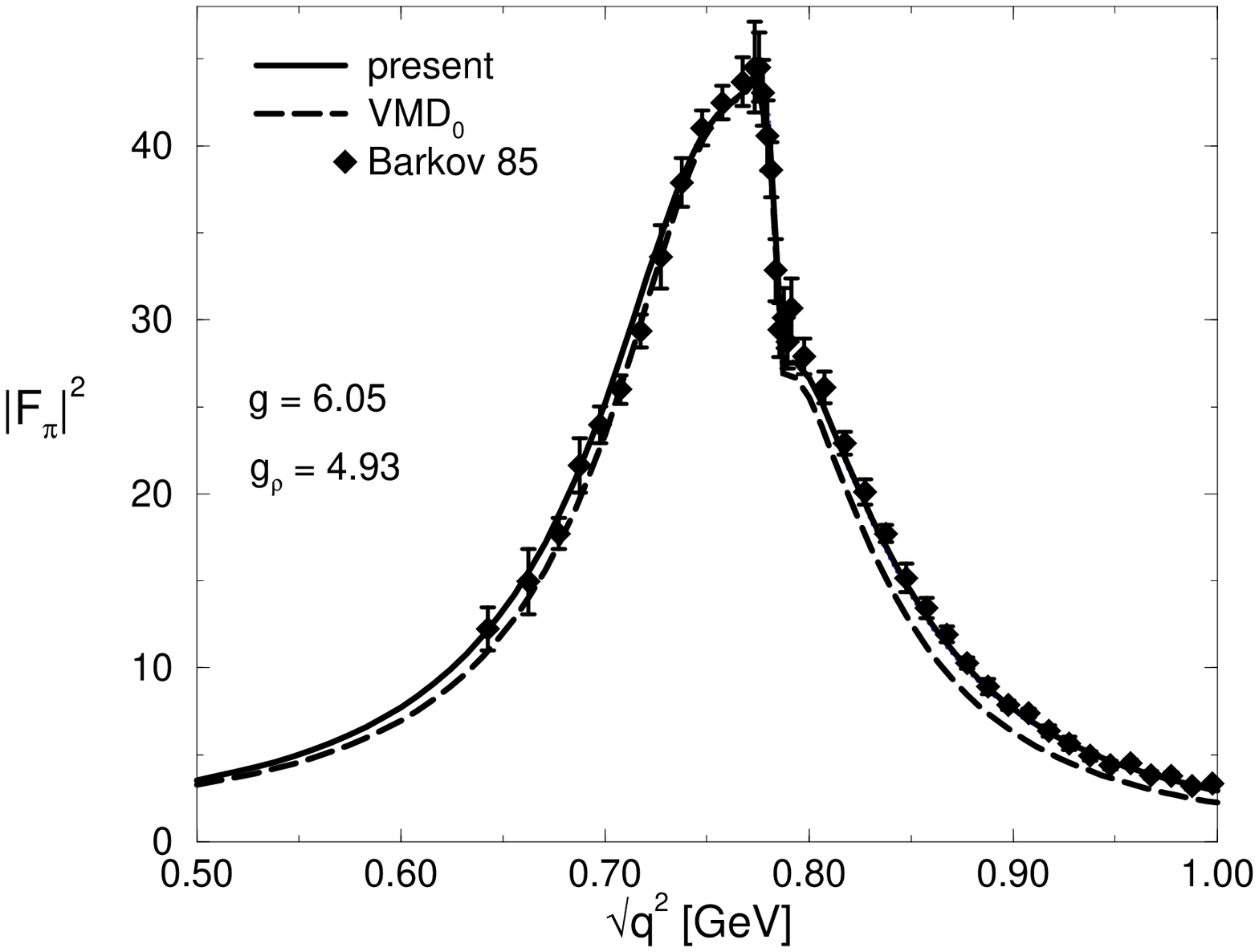}
\caption{\label{fig.8} }
\end{center} 
\end{figure}

Figure \ref{fig.8}: Pion form factor $F_\pi(q^2)$ in the region of timelike $q^2$. Data
from ref.\cite{14}. Solid curve: present result (eq.(\ref{5.28})) using $g=0.05$, $g_\rho=4.93$
and a $\rho \omega$ mixing parameter $z_{\rho \omega}=4.52 \cdot 10^{-3}
\GeV^2$. Dashed curve: standard VMD model result (eq.(\ref{5.29})).

\vspace{0.5 cm}

\begin{figure}
\begin{center}
\epsfysize9cm \leavevmode\epsfbox{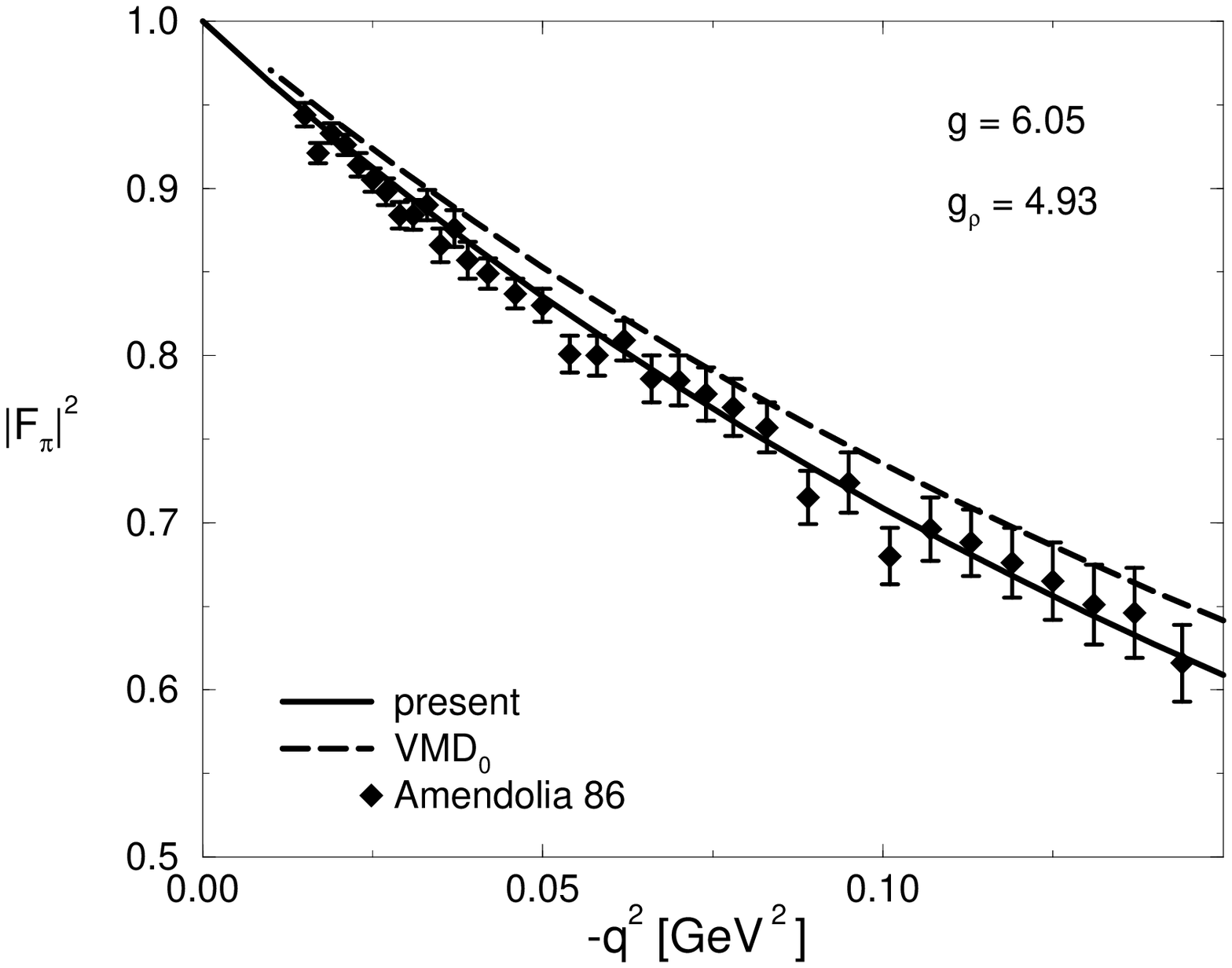}
\caption{\label{fig.9}}
\end{center}
\end{figure}

Figure \ref{fig.9}: Pion form factor $F_\pi(q^2)$ in the region of spacelike $q^2 <
0$. Data from ref.\cite{15}. Solid curve: present result
(eq.(\ref{5.28})) \vspace{0.5 cm}. Dashed curve: standard VMD model result (eq.(\ref{2.9})).

\vspace{0.5 cm}

\begin{figure}
\begin{center}
\epsfysize9cm \leavevmode\epsfbox{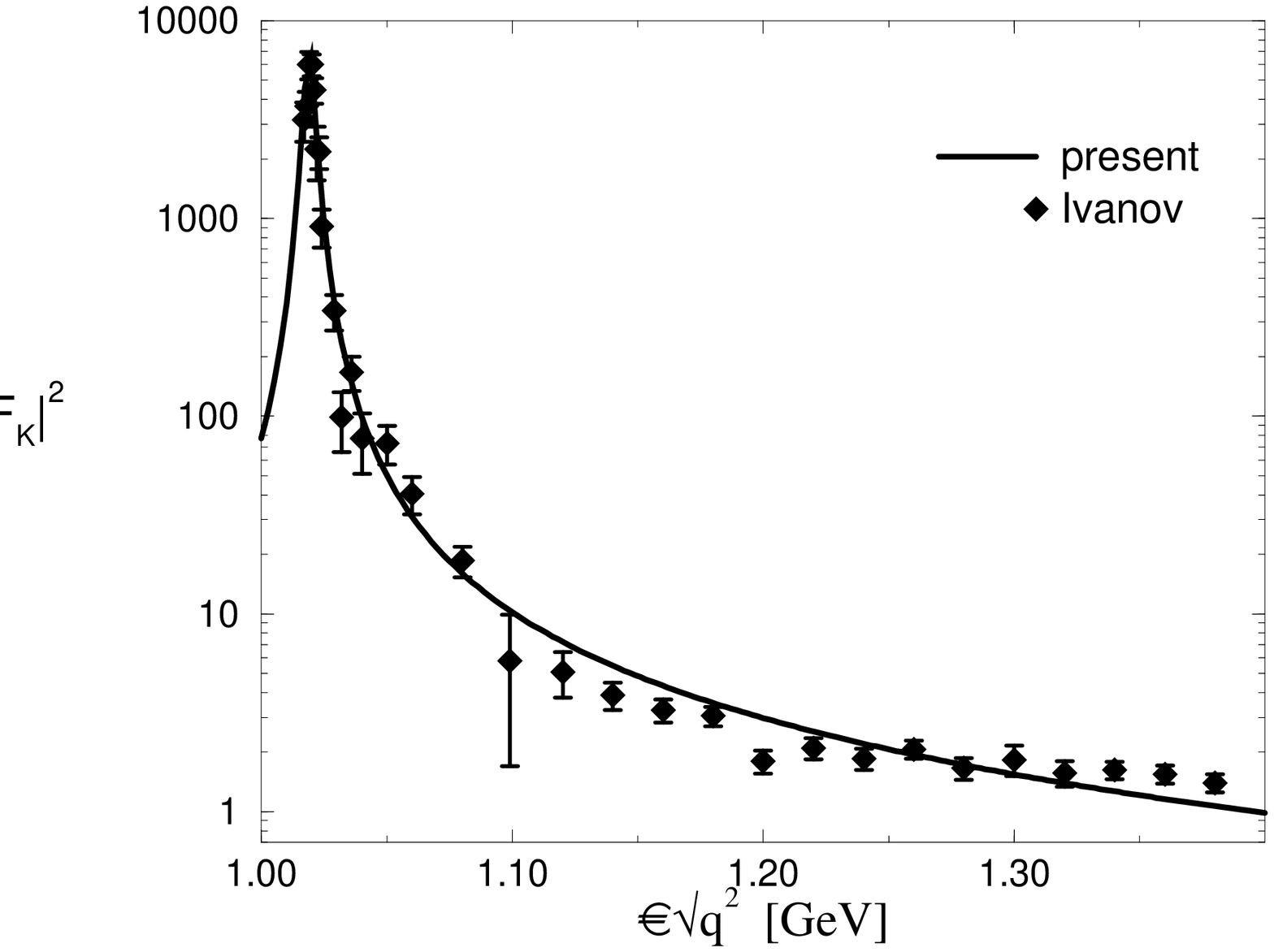}
\caption{\label{fig.10}}
\end{center} 
\end{figure}

Figure \ref{fig.10}: Kaon form factor $F_K(q^2)$ in the region of timelike $q^2>0$. The
solid curve is our result using eq.(\ref{5.43}). Data from ref.\cite{22}.

\vspace{0.5 cm}

\begin{figure}
\begin{center}
\epsfysize9cm \leavevmode\epsfbox{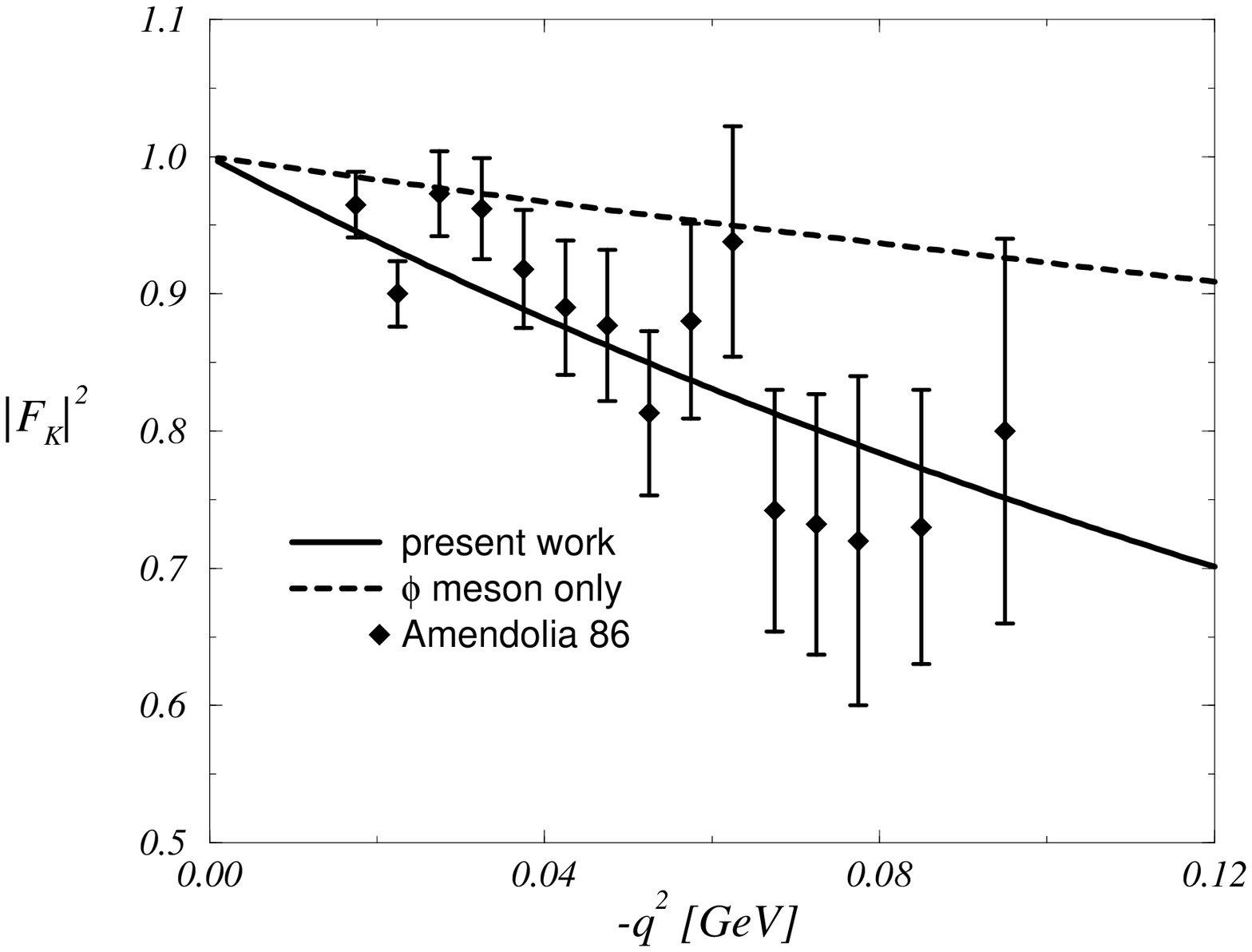}
\caption{\label{fig.11}}
\end{center}
\end{figure}

Figure \ref{fig.11}: Kaon form factor  $F_K(q^2)$ in the region of spacelike
$q^2<0$. Solid curve: our result using eq.(\ref{5.43}) including $\phi$, $\rho$ and
$\omega$ meson contributions. Dashed curve: $\phi$ contribution only. Data from
ref.\cite{18}.

\vspace{0.5 cm}

\begin{figure}
\begin{center}
\epsfysize9cm \leavevmode\epsfbox{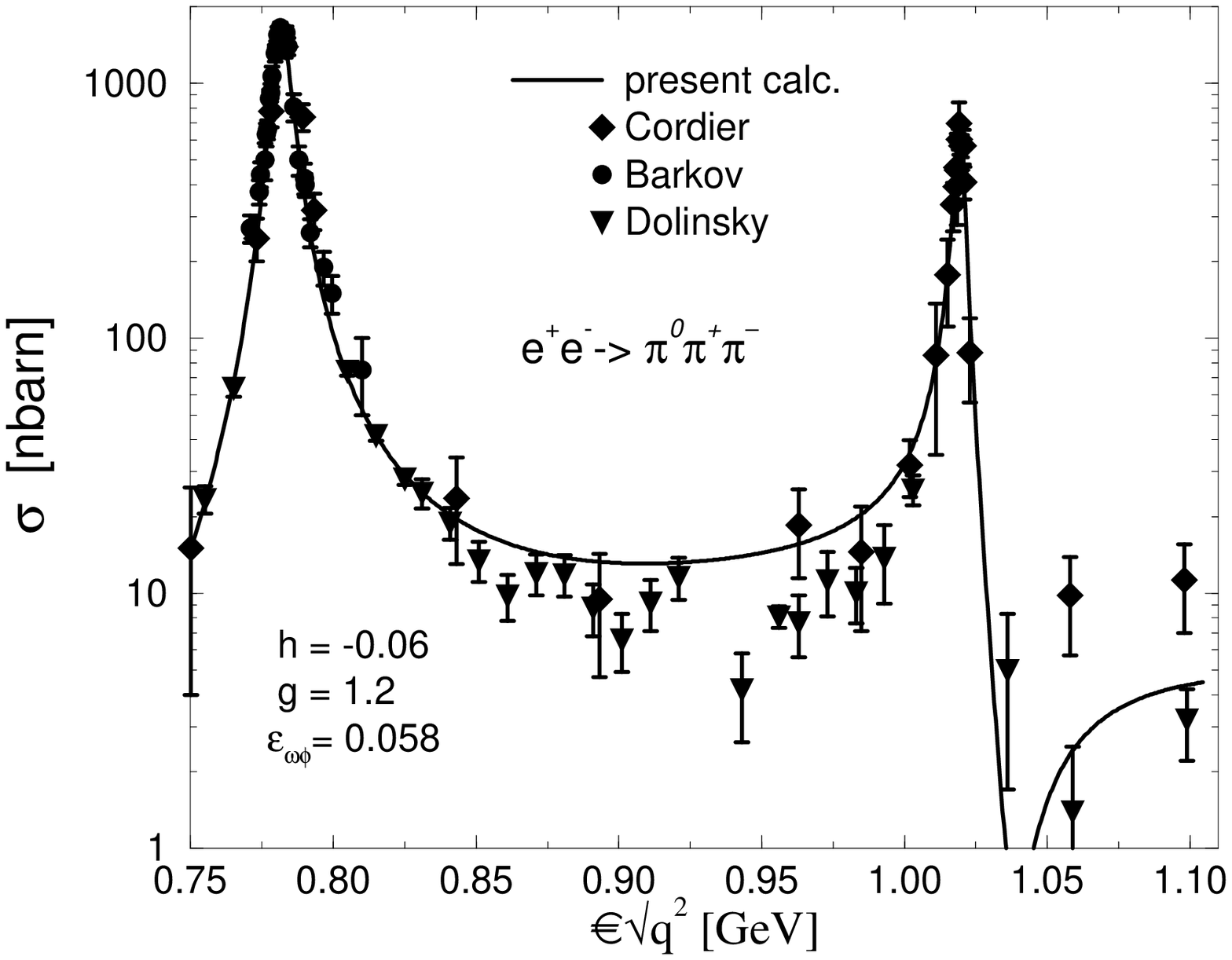}
\caption{\label{fig.12}}
\end{center}
\end{figure}

Figure \ref{fig.12}: Cross section for the reaction $e^+ e^- \to \pi^+ \pi^0
\pi^-$ as calculated in the
present work (solid line). See eq.(\ref{5.48}) and text for details. The data are taken
from refs.\cite{23,24,25}.

\vspace{0.5 cm}

\begin{figure}
\begin{center}  
\epsfysize5cm \leavevmode\epsfbox{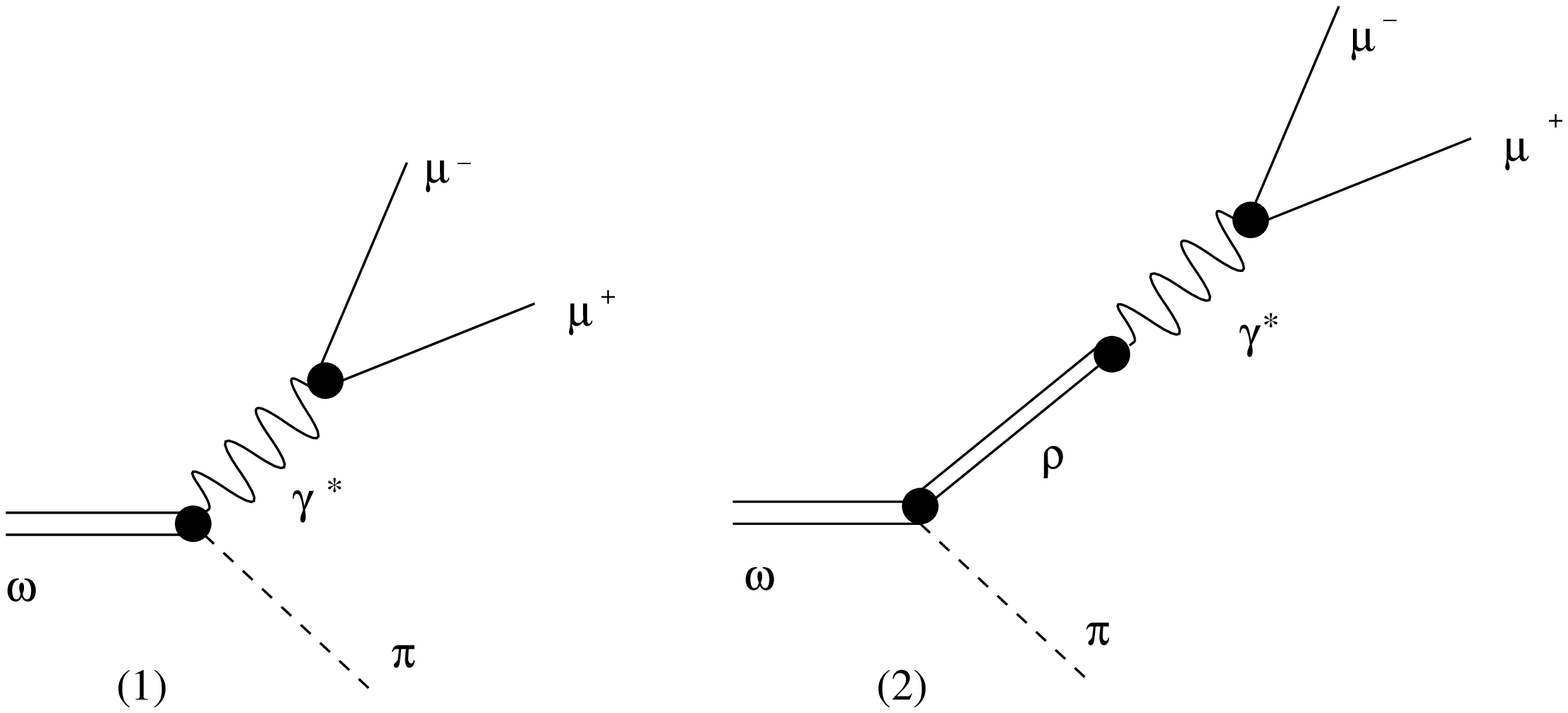}
\caption{\label{fig.13}}
\end{center} 
\end{figure}

Figure \ref{fig.13}: Basic mechanisms contributing to the Dalitz decay $
\omega \to \pi^0 \mu^+ \mu^-$: (1) direct $\omega \pi^0 \gamma$- coupling; (2)
two-step process via $\omega \to \pi^0 \rho$.

\vspace{0.5 cm}

\begin{figure}
 \begin{center}
\epsfysize10cm \leavevmode\epsfbox{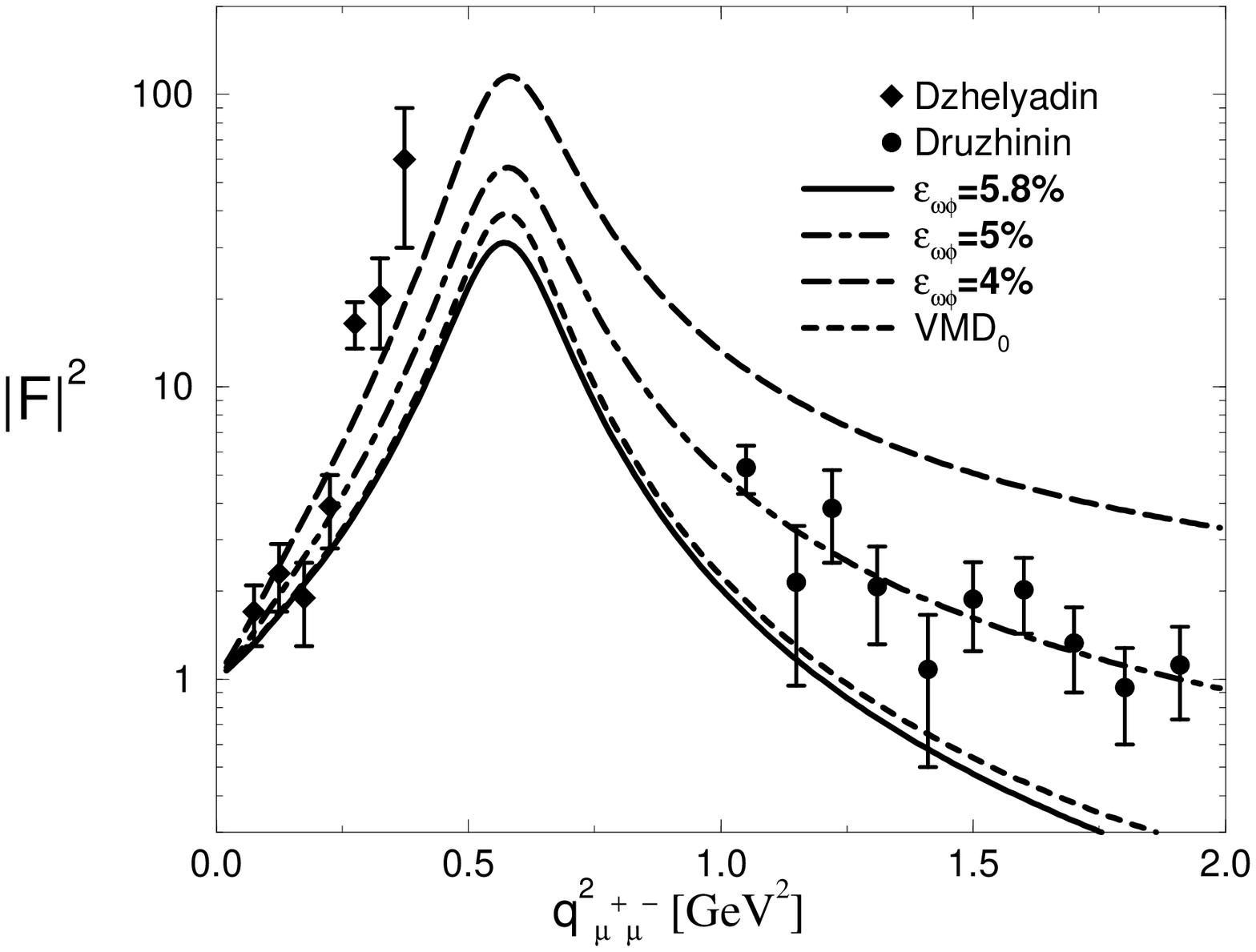}
\caption{\label{fig.14}}
\end{center}
\end{figure}

Figure \ref{fig.14}: Squared form factor $F(q^2)$ of the Dalitz decay $\omega \to
\pi^0 \mu^+ \mu^-$ as a function of the squared four-momentum of the muon
pair. Dashed, dotted and solid curves: our results with $\omega \phi$
mixing parameters $\epsilon_{\omega \phi}= 4 \cdot 10^{-2}$, $5 \cdot 10^{-2}$
and $5.8 \cdot 10^{-2}$ respectively. Short-dashed curve: standard
VMD-result. The experimental data are taken from ref.\cite{32,33}

\newpage

\end{document}